\pdfoutput=1
\documentclass[letterpaper,twocolumn,10pt]{article}
\usepackage[ruled]{algorithm2e} % For algorithms
\usepackage{balance}       % to better equalize the last page
\usepackage{graphics}      % for EPS, load graphicx instead
\usepackage{amsmath}
\usepackage{booktabs}
\usepackage{colortbl}
\usepackage{subfigure}
\usepackage{multirow}
\usepackage{xspace}
\usepackage{enumitem}
\usepackage{rotating}
\usepackage{caption}
\usepackage{epsf}
\usepackage{epsfig}
\usepackage{url}
\usepackage{pseudocode}
\usepackage{epstopdf}
\usepackage{ifpdf}
\usepackage{subfigure}
\usepackage{multirow}
\usepackage{tabularx}
\usepackage{diagbox}
\usepackage{color}
\usepackage{authblk}
\usepackage{tikz,hyphenat}
\usepackage{algpseudocode}
\usepackage{setspace}

\newcommand{\systemname}{LLAMA\xspace}
\newcommand{\systemnames}{LLAMA's\xspace}

\definecolor{Gray}{RGB}{199,201,203}

\newcommand{\parahead}[1]{\vspace{4pt plus 0pt minus 2pt}\noindent{\bfseries #1}}
\newcommand{\parabreak}{\vspace*{1.00ex minus 0.25ex}\noindent}

\usepackage{usenix2019_v3}

\makeatletter
\renewcommand\AB@affilsepx{, \protect\Affilfont}
\makeatother

\title{Pushing the Physical Limits of IoT Devices with Programmable Metasurfaces}
\author[1,2,*]{Lili Chen
\thanks{Work conducted on internship at Univ.~of Massachusetts Amherst.}}
\author[4]{Wenjun Hu}
\author[3]{Kyle Jamieson}
\author[2]{Xiaojiang Chen}
\author[2]{Dingyi Fang}
\author[1]{Jeremy Gummeson}
\affil[1]{Univ.~of Massachusetts Amherst}
\affil[2]{Northwest Univ.~(China)}
\affil[3]{Princeton Univ.}
\affil[4]{Yale Univ.}

\begin{document}
\maketitle

\begin{abstract}
\noindent{}Small, low-cost IoT devices are typically equipped with only a single, low-quality antenna, significantly limiting communication range and link quality. In particular, these antennas are typically linearly polarized and therefore susceptible to polarization mismatch, which can easily cause $10-15$~dB of link loss when communicating with such devices. In this work, we highlight this under-appreciated issue and propose the augmentation of IoT deployment environments with programmable, RF\hyp{}sensitive surfaces made of metamaterials.  Our smart metasurface mitigates polarization mismatch by rotating the polarization of signals that pass through or reflect from the surface. We integrate our metasurface into an IoT network as \systemname{}, a Low-power Lattice of Actuated Metasurface Antennas, designed for the pervasively used $2.4$~GHz ISM band. We optimize \systemnames{} metasurface design for both low transmission loss and low cost, to facilitate deployment at scale. We then build an end-to-end system that actuates the metasurface structure to optimize for link performance in real time. An empirical evaluation demonstrates gains in link power of up to 15~dB, and wireless capacity improvements of 100 and 180~Kbit\fshyp{}s\fshyp{}Hz in through\hyp{}surface and surface\hyp{}reflective scenarios, respectively, attributable to the polarization rotation properties of \systemnames{} metasurface.  
\end{abstract}
%-------------------------------------------------------------------------------

\section{Introduction}
\label{section: introduction}

Internet of Things (IoT) devices have achieved widespread adoption due to shrinking hardware costs and software management tools that ease installation by the end user. In recent years, a wide range of IoT devices have resulted in diverse systems including mobile devices such as smartwatches~\cite{nandakumar2016fingerio} and health trackers or statically deployed devices including sensors, cameras, voice assistants, and other appliance automation~\cite{chan2019detecting, chan2019contactless}. One key property these devices share is low\hyp{}cost hardware, in particular low\hyp{}cost radios, allowing for a minimal consumer price point. 
Such devices are typically deployed by non\hyp{}experts who understand neither their home's wireless environment nor the deployment considerations that govern wireless performance. Devices are typically deployed in a configuration that is well\hyp{}suited for a particular application or use\hyp{}case, but may not be the ideal placement in terms of communications performance. This combination of cheap hardware and non\hyp{}ideal network topology results in significant opportunities to improve wireless performance.

\begin{figure}[!t]
  \centering
 
  \includegraphics[width=0.4\textwidth]{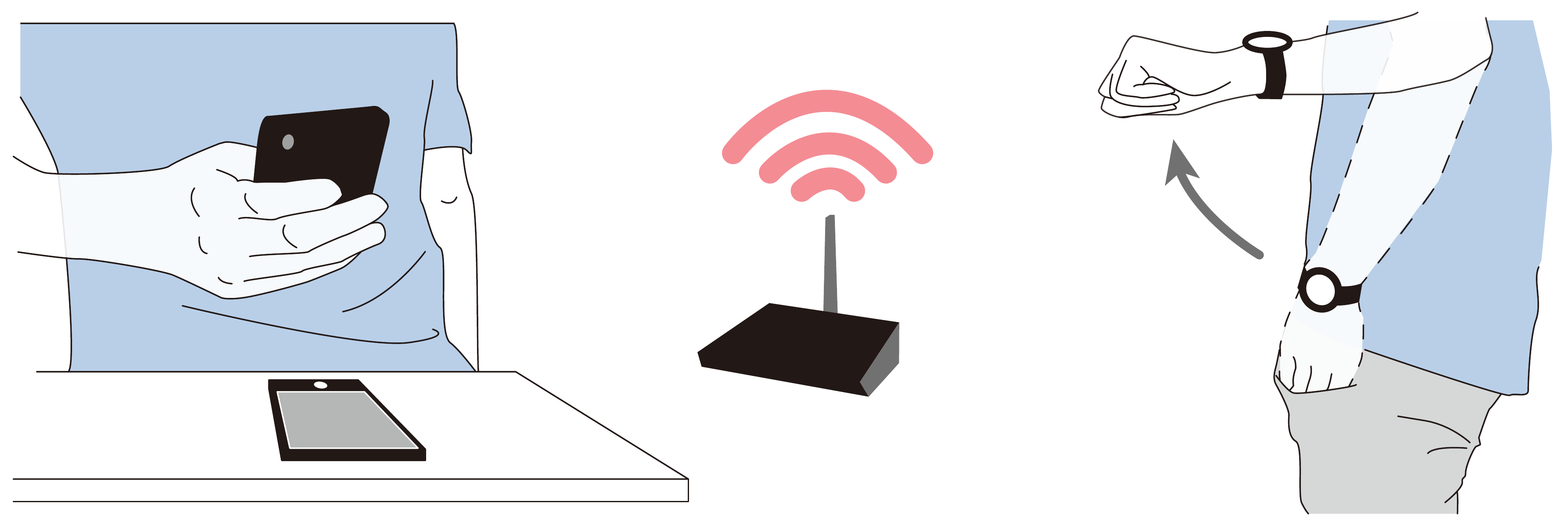}
  \caption{Low\hyp{}cost IoT devices and wearables suffer from polarization
  mismatch when their antennas become oriented perpendicularly with respect
  to an AP's antenna.} 
  \label{Mobile-devices}
\end{figure}

\begin{figure}[!t]
  \centering
  \subfigure[Wi-Fi communication (802.11g) between an AP~\cite{router} and a cheap ESP8266-based Arduino~\cite{esp8266}.]{
  \includegraphics[width=0.22\textwidth]{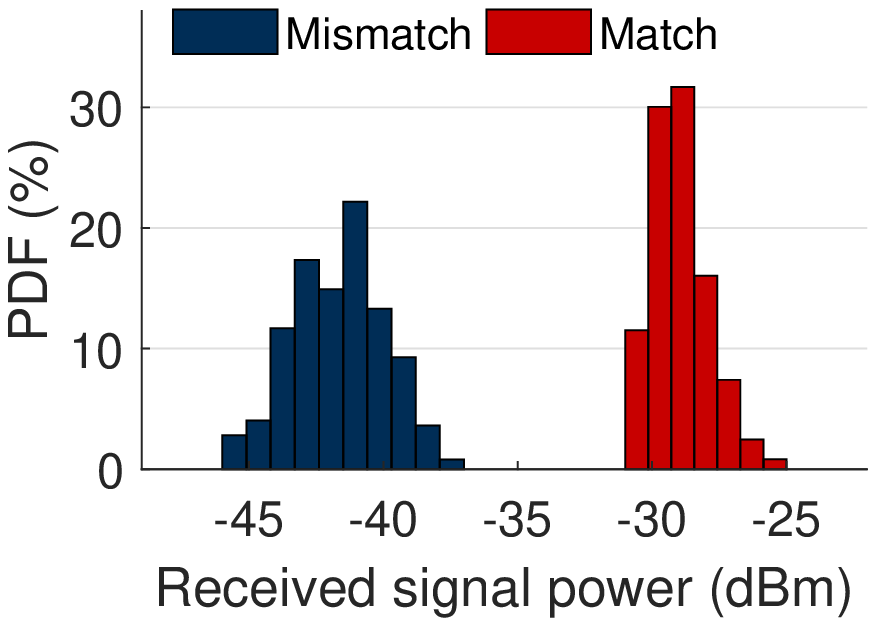}}
  \hspace{0.25cm}
  \subfigure[Bluetooth communication between a Huawei Watch~\cite{watch} and a Raspberry Pi 3~\cite{raspberrypi}. ]{
  \includegraphics[width=0.22\textwidth]{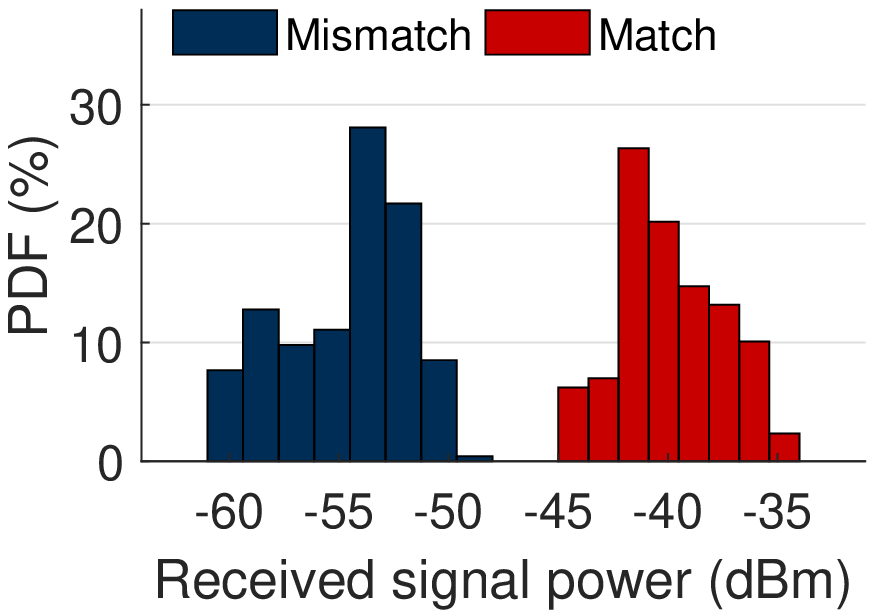}}
  %\vspace{-0.2cm}
  \caption{\emph{Impact of polarization mismatch.} Received signal power distributions for matching and mismatching antenna orientations between IoT transmitter\hyp{}receiver pairs. 
  Polarization mismatch significantly reduces link signal power.}
  \label{motivation}
\end{figure}

One source of performance degradation in such deployments is a significant power loss caused by a \textit{polarization mismatch}~\cite{bowers2015dynamic, okatan2009polarization} between a low\hyp{}cost dipole antenna on an IoT device and antennas on a Wi-Fi access point (AP)---in higher performance devices (\emph{i.e.}, mobile handsets) this loss is usually mitigated through the use of circularly polarized or dynamically\hyp{}switched linearly polarized antennas in different orientations. Low\hyp{}cost IoT devices instead use one cheap, linearly polarized dipole antenna that results in weak, fragile links between transmitters and  receivers. In addition to misaligned stationary devices, mobile devices such as wearables can suffer from dynamic antenna misalignment as a user swings their arm, for example, as Figure~\ref{Mobile-devices} illustrates.  This effect can be significant: microbenchmark experiments show that moving between orthogonal and aligned relative antenna polarization results in $\approx 10$~dB of power variation at the receiver, for both the low power Wi-Fi link between an Arduino equipped with an ESP8266 module~\cite{esp8266} and an 802.11g Wi-Fi AP~\cite{router} shown in Figure~\ref{motivation}~(a), and for a Bluetooth link between a smartwatch \cite{watch} and a Raspberry Pi 3~\cite{raspberrypi} shown in Figure~\ref{motivation}~(b).

In this work, we specifically investigate how to change the effective relative orientation of antennas at the communication endpoints, without hardware modifications to the endpoints themselves.  Achieving this objective would allow us to maintain a low bill of materials cost for IoT wireless communications components, and significantly improve the performance for existing off-the-shelf devices. Our approach to changing effective polarization alignment hinges instead, on changing the radio propagation environment itself, with a \textbf{\emph{Low-Power Lattice of Actuated Metasurface Antennas}} (\textit{\systemname}), a tunable smart surface made with inexpensive metamaterials~\cite{shrekenhamer2013liquid, shen2009broadband}.  LLAMA is deployed in the radio environment near the IoT endpoints, and is able to change the polarization of incident waves as they travel from sender to receiver.  As we show in this paper, the LLAMA substrate can be programmed dynamically to effect just the right amount of polarization rotation needed to help the ongoing communication between nearby endpoints. LLAMA follows the PRESS~\cite{welkie2017programmable} approach, which was the first to outline a vision of programmable radio environments along with a preliminary experimental setup that validates the feasibility of change the perceived channel at the communication endpoints. 

%Our system is inspired by a tunable polarization rotator design proposed for millimeter-wave ($10~GHz$) communication systems~\cite{wu2019tunable}, which relies on varying the bias voltage of integrated varactor diodes to implement two tunable, orthogonal phase shifters used for the purpose of controlling polarization.

Designing a metasurface in the $2.4$~GHz ISM band requires us to overcome significant challenges. First, longer wavelengths in the $2.4$~GHz band require larger and thicker metasurface substrates, which can attenuate the incident signal significantly. While an inefficient structure could rotate polarization, losses would dominate and the structure would attenuate the incident signal, hampering communication. Therefore, the metasurface structure needs to be optimized for low transmission loss.
Second, since we aim to realize pervasive deployments of these structures, we need to develop materials that are low cost, avoiding high performance but relatively expensive RF materials commonly used in other implementations of metasurfaces. %; for example, if Rogers 5880 was used as the substrate in our design, the resulting cost would be $\sim\$20000$~ for $0.25~m^2$ of surface area. 
%Second, low transmission loss through the metamaterial structure is needed, since the end goal is to provide overall signal improvement.  While an inefficient structure could rotate polarization, losses would dominate and the structure would attenuate the incident signal, hampering communication. 
Overcoming both of these challenges results in a pervasively deployable substrate that can compensate for losses between different endpoint pairs. 

%The relatively longer wavelength of the ISM band results in a design that is  ~($4\times$) larger than a metasurface designed for the~($10~GHz$) band; this precludes the use of high performance, low-loss substrate materials, i.e. Rogers 5880, as this would be cost prohibitive~(up to $\$20000$ for $0.25~m^2$ units of surface area). 

Indeed, naively replacing a high performance substrate (\emph{i.e.}, Rogers 5880~\cite{Rogers}) with a low-cost substrate (\emph{i.e.}, FR4~\cite{FR4}) results in higher transmission loss due to FR4's inherent physical properties; this in turn significantly attenuates power at the receiver, reducing link throughput and communication distance. To deal with this problem, we optimize the metasurface structure to ensure the overall system has both low transmission loss, as well as a scalable price point. Specifically, we choose a cheap material (FR4) as the substrate, use a minimum number of substrate layers for the required bandwidth, and minimize the thickness of each layer %$6.3~mm$~(scaled from $10~GHz$ structure) to $0.6~mm$
 to significantly reduce the losses associated with FR4. %One may ask: \emph{is there any drawback to such a minimal design?} The short answer is yes, as the main problem induced by this optimization is that it breaks the property of linear rotation across a wide communication bandwidth~($300~MHz$)%, because the transmission bandwidth varies linearly with the length of the transmission line~(thickness of the phase shifter). 
%~-- However, we observed that this has no impact on the $2.4~GHz$ ISM frequency band used by IoT devices, as the property of linearity \emph{is} maintained across the relatively narrow bandwidth ($100~MHz$). 
%Within this narrow bandwidth, the rotation angle is still linearly changes as the bias voltage of varactor diodes. Moreover, the fabrication cost of our structure is relative low since we utilized less amount of boards to act as tunable phase shifter and thus less number of diodes required.

To enable real-time polarization optimization, a receiver must report received power to a controller which in turn rotates polarization by modifying a pair of bias voltages. We provide a novel method to estimate the polarization rotation angle induced by the metasurface which can vary with link distance---understanding this mapping can enable rotation sensing and tracking.

\parahead{Contributions.} To summarize, \systemname is the first system that leverages an inexpensive RF substrate to optimize the radio environment in real time, thereby  avoiding signal losses caused by polarization mismatch, and thus enables higher quality communication links between IoT devices. In this work, we optimize a metasurface structure based on microwave attenuation theory and achieve comparable polarization tunability to a similar system that uses relatively expensive materials. We validate a proof-of-concept implementation of \systemname for both communication and sensing with comprehensive experiments. Our results show that \systemname enables polarization rotation within $3^\circ-45^\circ$, improves the signal strength by $15$~dB~(transmission) and $17$~dB~(reflection) with respect to mismatched antenna polarizations. \systemname also holds great potential to enhance sensing applications, as demonstrated in \S~\ref{evaluation: sensing}.

\section{A Polarizing View of Wireless}

Electromagnetic polarization describes the parametric trajectory of the electric and magnetic field vectors of a planar electromagnetic wave as it propagates through space. The polarization of RF wave propagation is a fundamental characteristic of wireless communication, but it has not received as much attention as issues like multipath fading and interference. An antenna constrains outgoing or incoming RF propagation to a particular plane. Therefore, communication is only possible if the signal propagation planes at both the transmitter and the receiver are well aligned. 

\begin{figure}[!t]
  \centering
  \vspace{0.15cm}
  \includegraphics[width=0.47\textwidth]{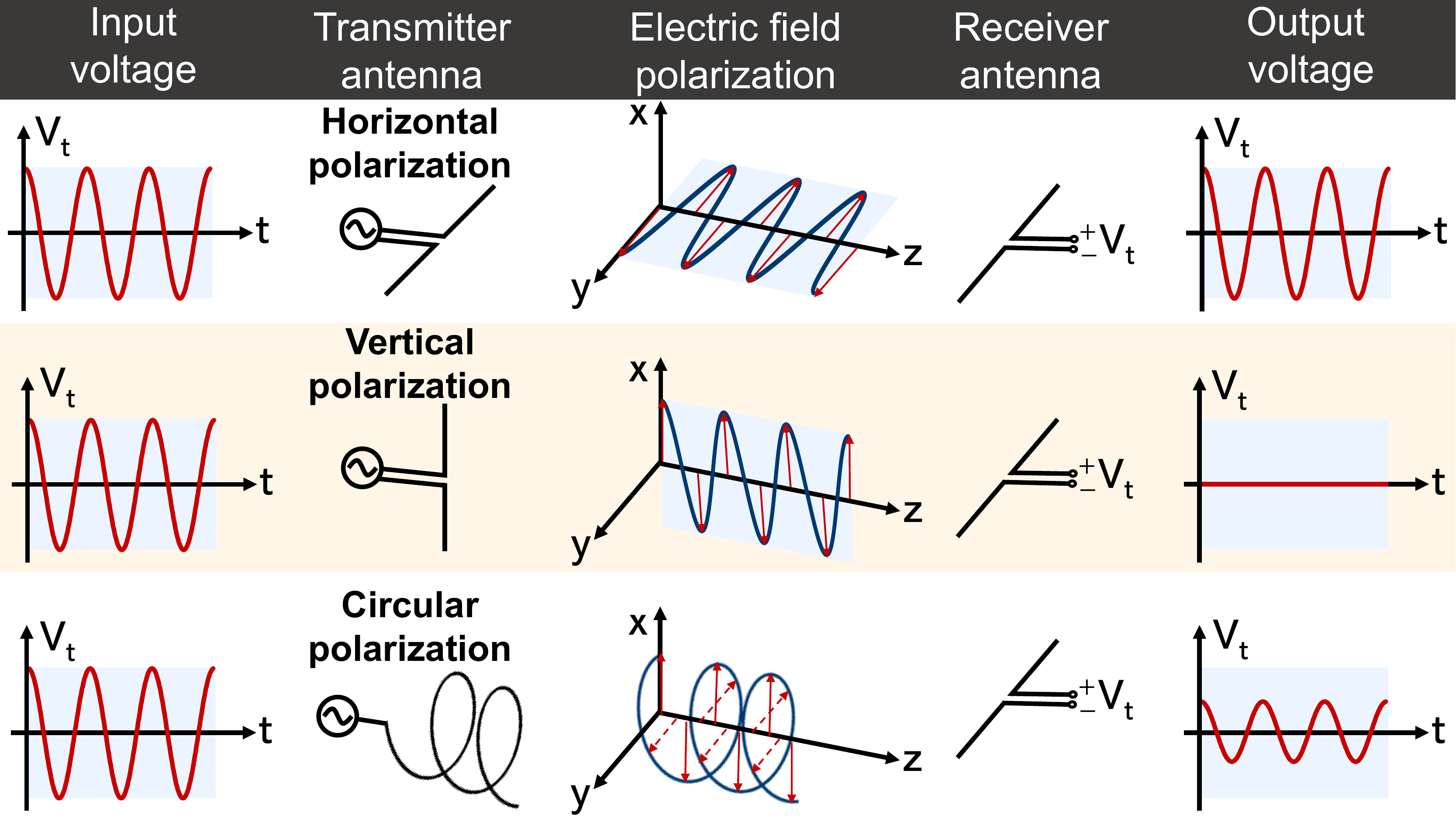}
  \caption{Signal transmission loss under different antenna polarization combinations between transmitter and receiver~\cite{bowers2015dynamic}.}
  \label{polarization}
\end{figure}

\parabreak{}\noindent\textbf{Polarization loss.}
One challenge in mobile wireless communication 
links is the significant power loss due to polarization mismatch. %Polarization mismatch is a phenomenon where additional coupling losses between the transmitter and receiver antennas occur on top of free-space propagation loss if the antennas are not of the same polarization and aligned in space.  
Three examples of various transmitter
antennas and their associated far-field electric fields are depicted in Figure~\ref{polarization}. If the signal is being received by a horizontally
polarized receiver antenna, it will be polarization matched to the transmitter antenna with horizontal linear polarization and the power received will primarily be a function of the transmit power and free space path loss. If the transmitter antenna is rotated in space,
the received signal will continue to degrade due to polarization
mismatch to the point where the very little signal is received
when the antennas are completely mismatched with orthogonal
polarizations. The signal loss is less when one of the antennas is circularly polarized~\footnote[1]{A theoretical $3$~dB degradation in coupling due to
polarization mismatch will also occur when one of the antennas
is circularly polarized while the other is linearly polarized.}.

As shown in Figure~\ref{motivation}, polarization mismatch can be debilitating for IoT devices. Higher performance devices such as mobile phones use switched antennas or circular polarized antennas to mitigate polarization mismatch, but low-cost IoT devices like smartwatches  typically have a single low-quality antenna.

\noindent\textbf{Correcting polarization mismatch.} 
Intuitively, polarization mismatch can be corrected by rotating the polarization of the signal before it arrives at the receiver. Here we show the mathematical foundation of polarization rotation.

In general, the polarization state of radio waves can be described by a $2\times 1$ Jones vector $J$. Consider a plane perpendicular to the direction of signal propagation. Any polarization state can be represented by two orthogonal components in that plane~(\emph{i.e.}, projected onto the $X$ and $Y$ axes) with different amplitude and phase. The \emph{Jones vector} is~\cite{kress2000digital}: 
\begin{equation}
J={\left[ \begin{array}{c}
a^{x}\\
a^{y}e^{j\pi/2} 
\end{array}
\right ]},
\end{equation}
where $a^{x}$ and $a^{y}$ represent the $X$ and $Y$ polarized signal components respectively.

When a manipulation surface is aligned with the x-y  coordinate axis, the \emph{Jones matrix} is defined as~\cite{kress2000digital}:
\begin{equation}
M=e^{j\alpha}{\left[ \begin{array}{cc}
1 & 0\\
0 & e^{j\pi/2} 
\end{array}
\right ]},
\end{equation}
where $\alpha$ is a phase delay between the $X$ and $Y$ axes.
If the surface is rotated counterclockwise by a degree of $\theta$, the Jones matrix becomes~\cite{kress2000digital}:
\begin{equation}\label{rotation-matrix}
M_{\theta}=R(\theta)MR(\theta)^{T}, ~~~~~~~ R(\theta)={\left[ \begin{array}{cc}
\cos\theta & -\sin\theta\\
\sin\theta & \cos\theta 
\end{array}
\right ]},
\end{equation}
where $R(\theta)$ is a rotation matrix.

In systems with multiple layers of polarization manipulation surface between the incident wave and outgoing wave, the outgoing Jones vector $J_{out}$ is obtained by multiplying the Jones vector of incident wave with the Jones matrix of each surface layer~\cite{kress2000digital}:
\begin{equation}
J_{out}=M_{N}...M_{2}M_{1}J_{in},
\end{equation}
where $M_{N}$ is a $2\times 2$ Jones matrix, representing the $N^{\text{th}}$ surface.

%\begin{figure}[!t]
%  \centering
%  \includegraphics[width=0.3\textwidth]{figure/surface.pdf}
%  \caption{Jones matrix.}
%  \label{polarization}
%\end{figure}

\begin{figure}[!t]
  \centering
  \includegraphics[width=0.47\textwidth]{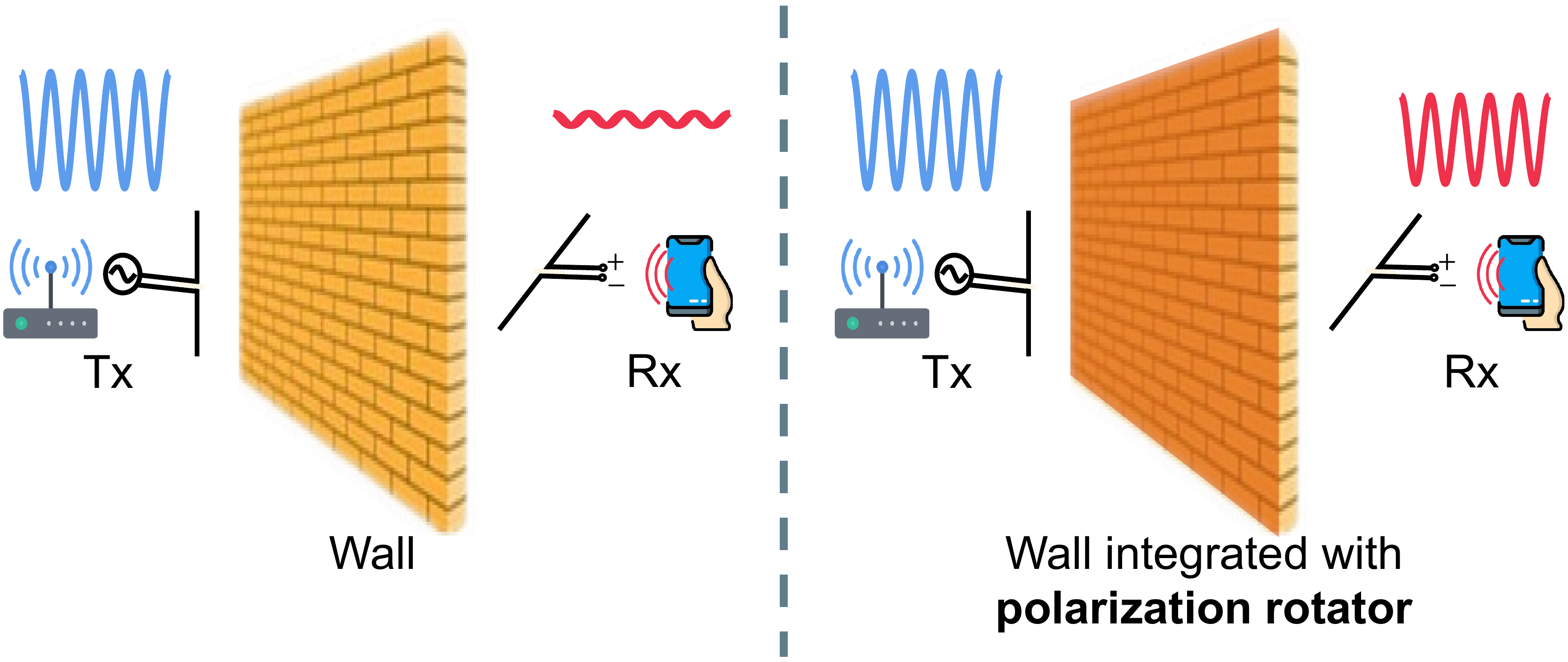}
  \caption{Wireless communication system without/with polarization rotator between mismatched transmitter~(Tx) and receiver~(Rx).}
  \label{overview}
\end{figure}

\section{System Design}

In this section, we introduce the \systemname architecture (\S~\ref{section:Metasurface-overview}) and illustrate the properties of the metasurface hardware with HFSS simulation results~(\S~\ref{section:Metasurface-architecture}). Next, we illustrate our approach towards actuating polarization angle in real-time by manipulating bias voltages for the polarization rotator~(in \S~\ref{section:bias-voltage-control}). While the primary goal of \systemname is to enhance signal quality between two endpoints, it can also be used for orientation sensing, which requires estimating relative polarization rotation between endpoint; \S~\ref{section:mapping} describes a technique for this purpose. %used to obtain a unique mapping relationship between the polarization rotation angle and received signal power variation to understand the relative endpoints orientation at arbitrary transmitter-receiver distance for sensing purposes.

\subsection{System Overview}
\label{section:Metasurface-overview}

\systemname is a low power system that is designed to reduce significant wireless signal loss caused by polarization mismatch between the transmitter and receiver. As shown in Figure~\ref{overview}, the signal from a mismatched transmitter arrives at the receiver with a lower loss when the intermediate wall includes a polarization rotator. \systemname has the ability to improve the communication quality and extend the sensing range in the widely used ISM frequency band.

\begin{figure}[!t]
  \centering
  \includegraphics[width=0.47\textwidth]{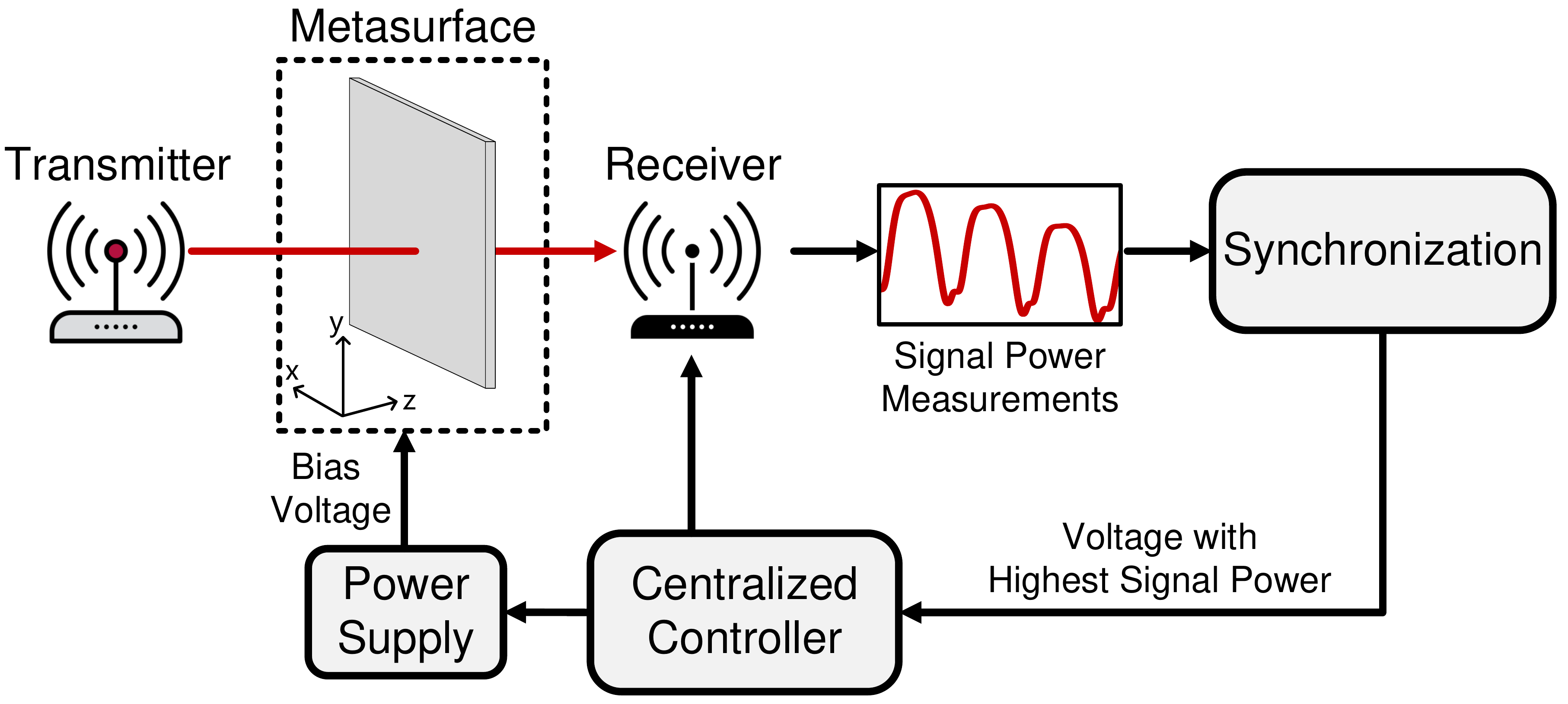}
  \caption{\systemname{} system architecture, showing transmitter
  and receiver endpoints, the \systemname{} substrate
  deployed in the environment, and control signal flow.}
  \label{architecture}
\end{figure}

An overview of our system architecture is depicted in Figure~\ref{architecture} and consists of these four elements:

\noindent\textbf{Metasurface.} The metasurface used in \systemname is a polarization rotator implemented using a low-cost FR4 substrate; the polarization rotator is tunable and uses biasing voltages within phase shifters in both the $X$ and $Y$ axes to define a rotation angle. The metasurface is deployed in a structural element (\emph{i.e.}, wall) and influences wireless signals that reflect from or propagate through the metasurface.

\noindent\textbf{Centralized controller.} A centralized controller observes the power measured at a receiver and uses a search algorithm to determine a set of bias voltages that maximize received signal power by finding the optimal rotation angle that achieves a polarization match between the antennas at the endpoints.

\noindent\textbf{Power supply.} The bias voltages used to tune the metasurface are set with a programmable DC power supply. By synchronizing the power supply output with the receiver, we can manipulate the polarization rotator with an optimal rotation angle in real time, with a rotation that maximizes signal power. Two bias voltages are needed for the phase shifters in the $X$ and $Y$ axes; while bias voltages as high as $30$~V are needed, the metasurface draws only $15$~nA of current. In future implementations, a circuit that generates these bias voltages could be integrated directly on the metasurface.

\noindent\textbf{Endpoints.} The endpoint receiver reports its received signal strength to the controller, which then determines how to actuate the metasurface by manipulating the two bias voltages.

% \begin{figure}[!t]
%   \centering
%   \subfigure[Structure scaled from~\cite{wu2019tunable}.\blue{$100$ USD per unit.}]{
%   \includegraphics[width=0.225\textwidth]{figure/original_structure.pdf}}
%   \hspace{0.1cm}
%   \subfigure[Optimized structure of \systemname.\blue{$5$ USD per unit.}]{
%   \includegraphics[width=0.225\textwidth]{figure/optimized_structure.pdf}}
%   \subfigure[Unit dimension of metasurface in \systemname.]{
%   \includegraphics[width=0.47\textwidth]{figure/unit_pattern.pdf}}
%   \caption{Polarization rotator structure at $2.4~GHz$ ISM frequency band. The rotator consists of a tunable birefringent structure (green) placed between two quarter wave plates~(blue). The quarter wave plates are rotated by $+45^{\circ}$ and $-45^{\circ}$ with respect to the birefringent structure, which makes the phase delays for two orthogonal polarizations differ by $90^{\circ}$. The tunable birefringent structure acts as a phase shifter that supports various polarization rotation degrees by manipulating its bias voltage of $X$ and $Y$ axis. The metallic patters~(orange) plated on the substrate boards~(green and blue) act as admittance.}
%   \label{rotator-submodule}
% \end{figure}

\subsection{Metasurface Architecture}\label{section:Metasurface-architecture}
Tunable metasurfaces are implemented as layered structures that consist of copper patterns printed on controlled dielectric substrates; these layers perform different functions that reflect, bias, or guide EM waves. A transmissive metasurface uses a biasing network sandwiched between or adjacent to waveguide layer(s) to modify the transmissive signal properties in a controlled manner. In contrast, a reflective metasurface uses a metallic plane as one of its layers, where the signal passes through a wave guide layer and a biasing layer, and reflects from the metallic plane in the same relative direction with a different angle of departure. Depending on the bias voltages used \systemname can operate in either a transmissive or reflective mode.

\noindent\textbf{Cost-effective metasurface design.}
%Our simulated polarization rotator was inspired by a $10$~GHz design~\cite{wu2019tunable} only demonstrated in a controlled environment with signal generators and horn antennas; translating this and similar designs into an end-to-end system that  demonstrates control over signal properties (e.g., polarization rotation angle) in realistic environments is the goal of our low-cost metasurface architecture.
Our design was inspired by a $10$~GHz design~\cite{wu2019tunable}, and we calculate the correct geometries of circuit elements for $2.4$~GHz instead based on the impedance matching. 
  
  To achieve polarization rotation for both x-polarized and y-polarized waves, we construct a polarization rotator consisting of a tunable birefringent structure~(BFS) placed between two quarter wave plates~(QWP~\cite{davis2000two}. The QWPs are rotated by $+45^{\circ}$ and $-45^{\circ}$ with respect to the BFS, which causes the phase delays for two orthogonal polarizations differ by $90^{\circ}$. The Jones matrices of the two QWPs can be expressed as:

\begin{equation}
Q_{+45^{\circ}}=e^{j\alpha}R(+45^{\circ}){\left[ \begin{array}{cc}
1 & 0\\
0 & e^{j\pi/2} 
\end{array}
\right ]}R(+45^{\circ}),
\end{equation}
\begin{equation}
Q_{-45^{\circ}}=e^{j\alpha}R(-45^{\circ}){\left[ \begin{array}{cc}
1 & 0\\
0 & e^{j\pi/2} 
\end{array}
\right ]}R(-45^{\circ}).
\end{equation}
The tunable BFS is a transmissive metasurface that can rotate the polarization of the $X$ and $Y$ axes, independently. The Jones matrix of the BFS is:
\begin{equation}
B=e^{j\beta}{\left[ \begin{array}{cc}
1 & 0\\
0 & e^{j\delta} 
\end{array}
\right ]},
\end{equation}
where $\beta$ is the transmission phase as the signal passes through the BFS, irrespective of initial polarization orientation, while $\delta$ represents the transmission phase difference between the $X$ and $Y$ polarizations, which can be adjusted by manipulating the biasing voltages of the $X$ and $Y$ axes as shown in Figure~\ref{architecture}.
The entire Jones matrix of the polarization rotator is:
\begin{equation}
\begin{aligned}	
P&=Q_{+45^{\circ}}BQ_{-45^{\circ}}\\
&=e^{j(\alpha+(\pi/2)+\beta+(\delta/2))}{\left[ \begin{array}{cc}
\cos(\delta/2) & -\sin(\delta/2)\\
\sin(\delta/2) & \cos(\delta/2) 
\end{array}
\right ]}.
\end{aligned}
\end{equation}
In summary, the proposed structure can rotate the polarization of a wave by $\delta/2$ rotation degrees, according to the rotation matrix presented in Equation~(\ref{rotation-matrix}).

\begin{figure}[!t]
  \centering
  \subfigure[Metasurface topology of \systemname.]{
  \includegraphics[width=0.47\textwidth]{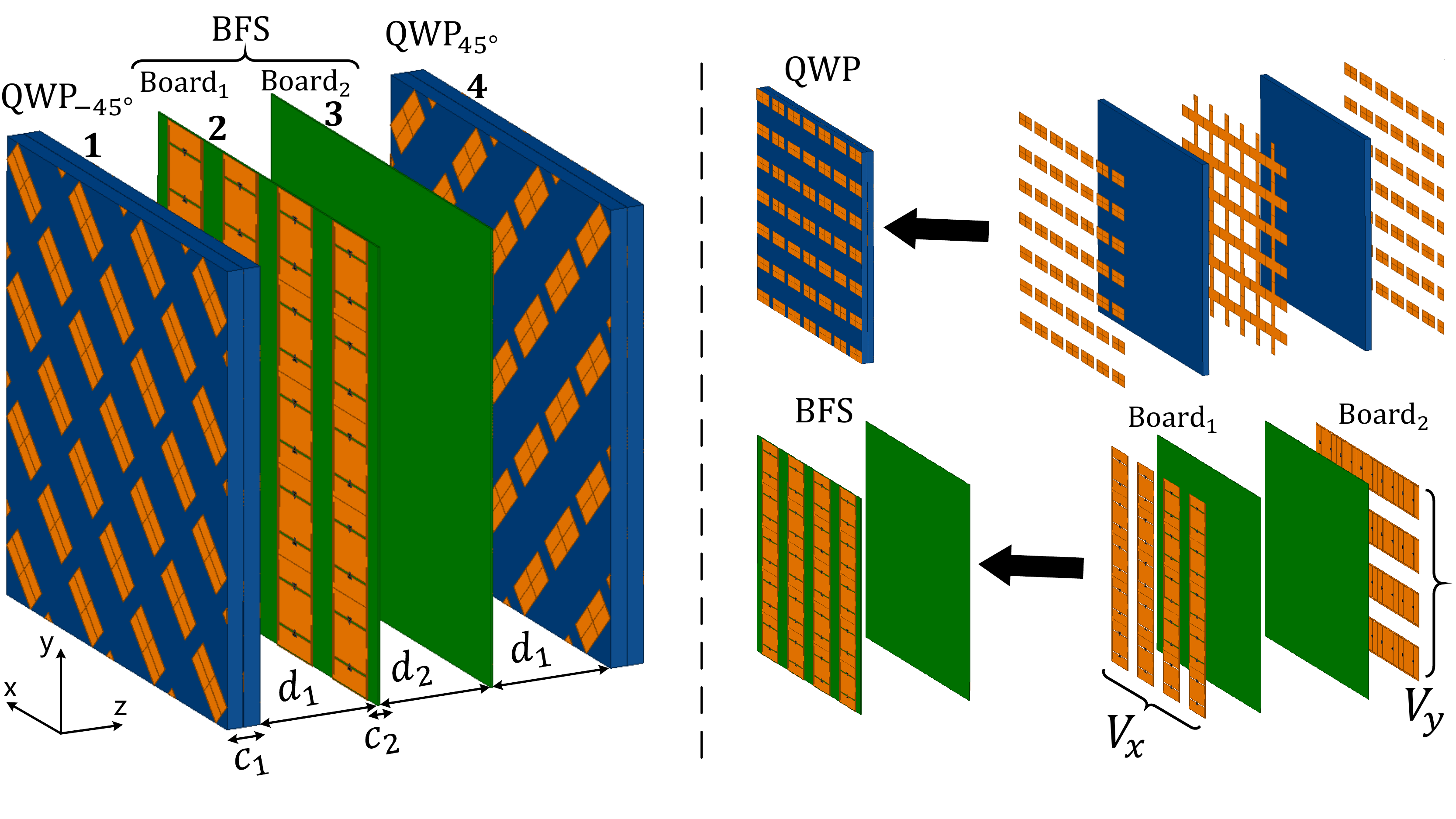}\label{metasurface:layers}}
  \subfigure[Unit dimension of metasurface in \systemname. The black elements in BFS represent the varactor diodes.]{
  \includegraphics[width=0.47\textwidth]{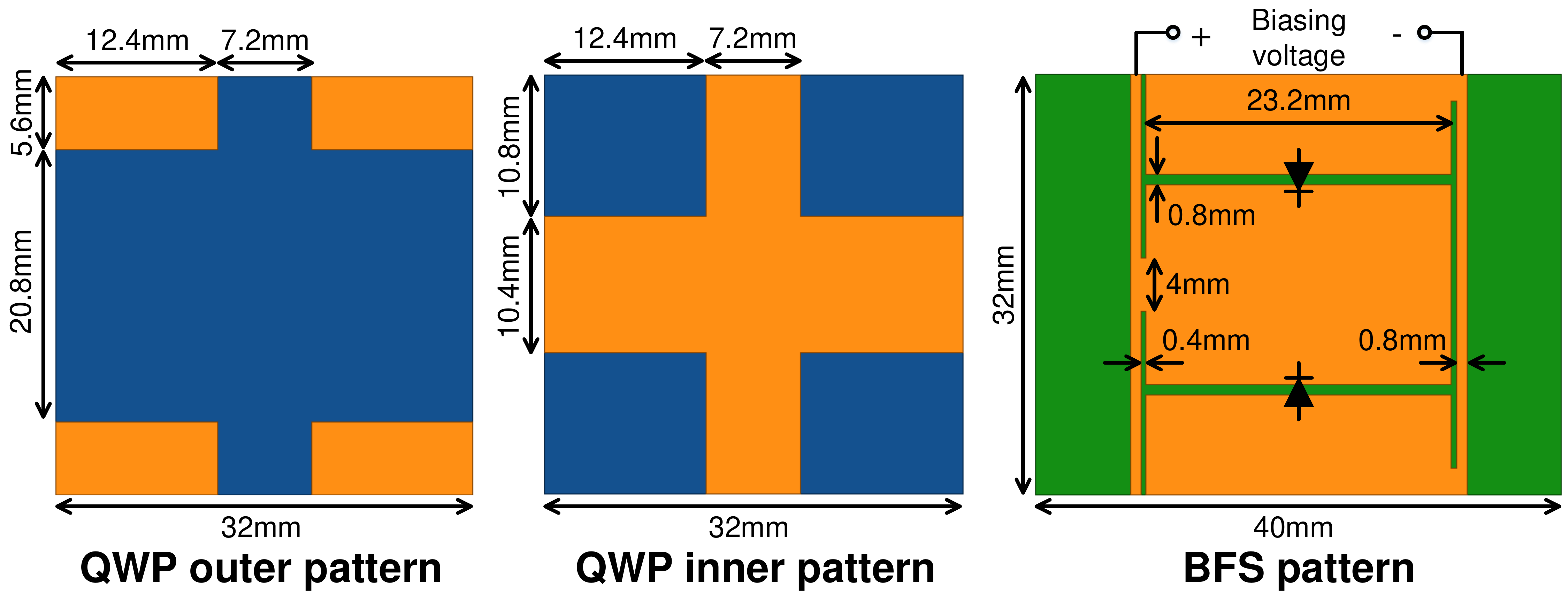}}
  %\vspace{-0.2cm}
  \caption{Polarization rotator structure designed for the $2.4$~GHz ISM band. The distances between adjacent boards are $d_1=7$~mm and $d_2=11$~mm, the FR4 substrate thicknesses of QWP and BFS are $c_1=12$~mm and $c_2=0.6$~mm.}
  \label{rotator-submodule}
\end{figure}

In Figure~\ref{rotator-submodule}~(a) we show the microstrip geometries used in our metasurface design. The metasurface consists of a tunable BFS (layers 2 and 3 in Fig.~\ref{metasurface:layers}) placed between two QWPs (layers 1 and 4 in Fig.~\ref{metasurface:layers}) rotated $45^{\circ}$ with respect to the BFS. 
  The BFS includes two birefringent boards rotated by $90^{\circ}$ with respect to each other, each board acts as a phase shifter that supports different polarization rotations by manipulating the bias voltages~($V_x$ and $V_y$) of integrated varactor diodes~(black) along $X$ and $Y$ axes. The metallic patterns~(orange) plated on the substrate boards~(green and blue) act as admittance components.

\begin{figure}[!t]
  \centering
  \includegraphics[width=0.35\textwidth]{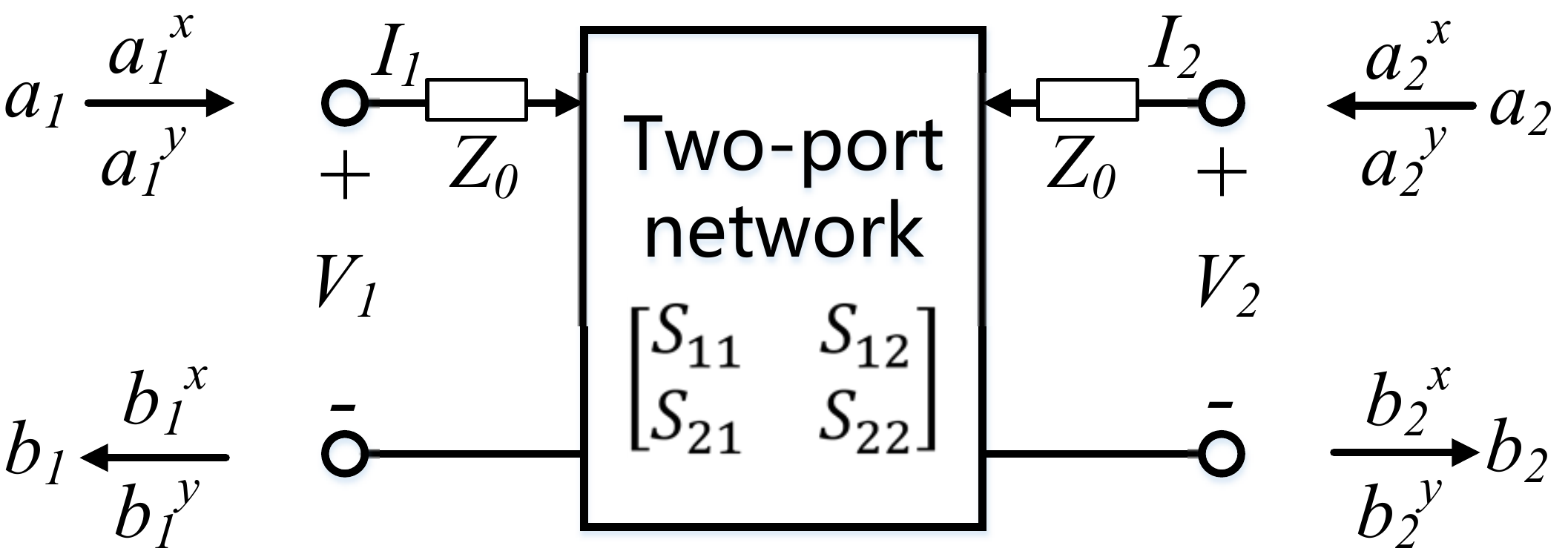}
  \caption{The scattering parameters of metasurface can be measured according to the matched impedance of the input and output ports~\cite{rao2015microwave}.}
  \label{2-port}
\end{figure}

% \begin{figure*}[!t]
%   \centering
%   \subfigure[x-polarized incident wave.]{
%   \includegraphics[width=0.47\textwidth]{figure/x_polarized_efficiency.eps}}
%   \hspace{0.5cm}
%   \subfigure[y-polarized incident wave.]{
%   \includegraphics[width=0.47\textwidth]{figure/y_polarized_efficiency.eps}}
%   \caption{$S_{21}$ efficiencies for x- and y-polarized incident waves under different voltage combinations of $X$ and $Y$ axes. The results indicate that the polarization can be controlled by changing the biasing voltages of phase shifter.}
%   \label{polarization_results}
% \end{figure*}
We next consider the transmission efficiency~($S_{21}$) of $X$ and $Y$ axis polarized signals, as that is among the most important performance metric of the overall metasurface design. Higher transmission efficiency indicates better performance. %To better understand the transmission efficiency, here we provide a brief introduction to the scattering matrix $S$. 
For a two-port network as shown in Figure~\ref{2-port}, the amplitude~(normalized voltage) of incoming waves ($a_1$ and $a_2$) and the outgoing waves ($b_1$ and $b_2$) are given by~\cite{rao2015microwave}:
\begin{equation}\label{wave}
\left\{
             \begin{array}{lr}
            a_{1}=\frac{V_{1}+Z_{0}I_{1}}{2\sqrt{Z_{0}}}& \vspace{1ex} \\
          
            a_{2}=\frac{V_{2}+Z_{0}I_{2}}{2\sqrt{Z_{0}}}&
             \end{array}
\right.,~~~~~~~~~~
\left\{
             \begin{array}{lr}
            b_{1}=\frac{V_{1}-Z_{0}I_{1}}{2\sqrt{Z_{0}}}& \vspace{1ex} \\
             b_{2}=\frac{V_{2}-Z_{0}I_{2}}{2\sqrt{Z_{0}}} &  
             \end{array}
\right.,
\end{equation}
where $V_1$ and $V_2$ are the normalized voltage of port $1$ and port $2$, $I_1$ and $I_2$ are the normalized current of port $1$ and port $2$, $Z_0$ is the matched impedance.
The scattering matrix $S$ relates the incoming waves to the outgoing waves as~\cite{rao2015microwave}: 
\begin{equation}\label{scatter-matrix}
{\left[ \begin{array}{c}
b_{1}\\
b_{2} 
\end{array}
\right ]}
={\left[ \begin{array}{cc}
S_{11}& S_{12} \\
S_{21}& S_{22}
\end{array}
\right ]} \times 
{\left[ \begin{array}{c}
a_{1}\\
a_{2} 
\end{array}
\right ]},~
S_{ij}=\frac{b_{i}}{a_{j}}|a_{k}=0~\forall~k\neq j .
\end{equation}
$S_{11}$ and $S_{22}$ are reflection coefficients,
$S_{21}$ and $S_{12}$ are transmission coefficients.
Then the transmission efficiency can be calculated according to the following equation~\cite{rao2015microwave}:
\begin{equation}\label{eff}
eff=\left\{
             \begin{array}{lr}
             |S_{21}^{xx}|^{2}+|S_{21}^{yx}|^{2} ~,~~for~~ x-polarized~~ wave& \\[8pt]
             |S_{21}^{xy}|^{2}+|S_{21}^{yy}|^{2} ~,~~for~~ y-polarized~~ wave &
             \end{array}
\right.,
\end{equation}
where $S_{21}^{yx}$ is obtained from the x-polarized component of incoming wave $a_1^{x}$ and the y-polarized component of outgoing wave $b_2^{y}$.

\begin{figure*}[!t]
  \centering
  \begin{minipage}[c]{0.31\textwidth}
  \includegraphics[width=1\textwidth]{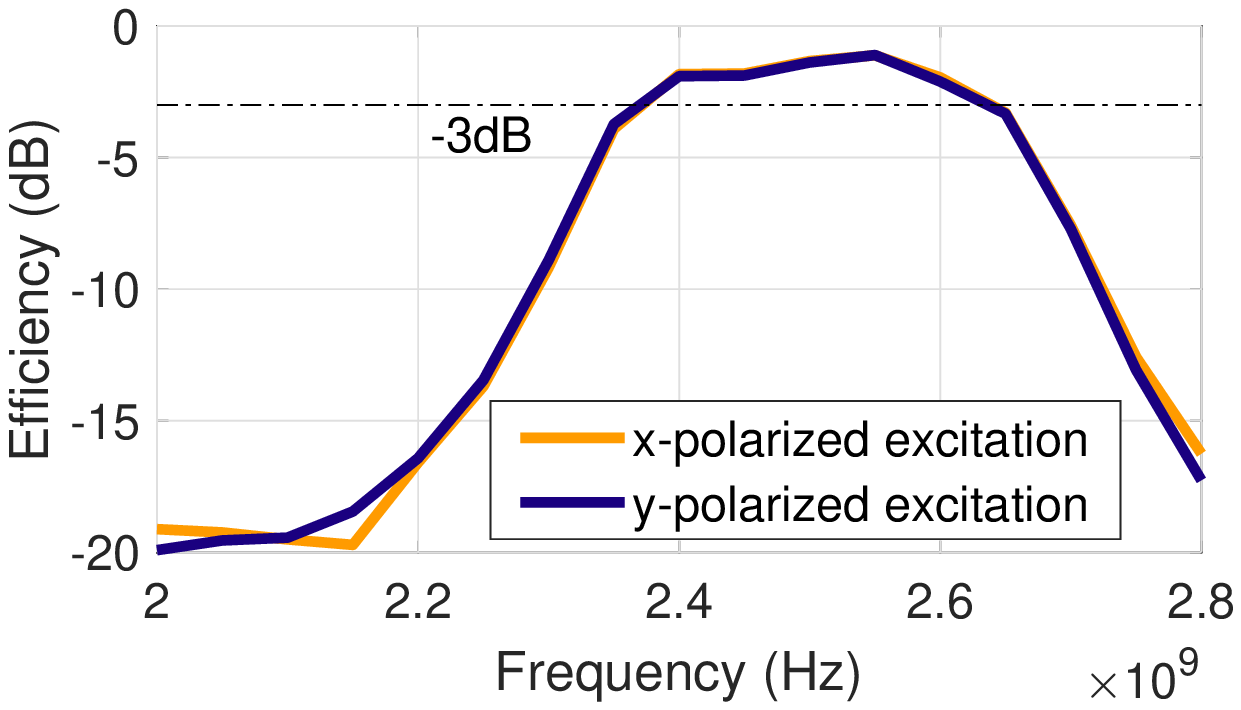}
  \caption{$S_{21}$ efficiency of cascaded polarization rotator layers using Rogers 5880 substrate~(loss tangent is $0.0009$).}
  \label{Rogers_efficiency}
  \end{minipage}
  \hspace{0.25cm}
  \begin{minipage}[c]{0.31\textwidth}
  \includegraphics[width=1\textwidth]{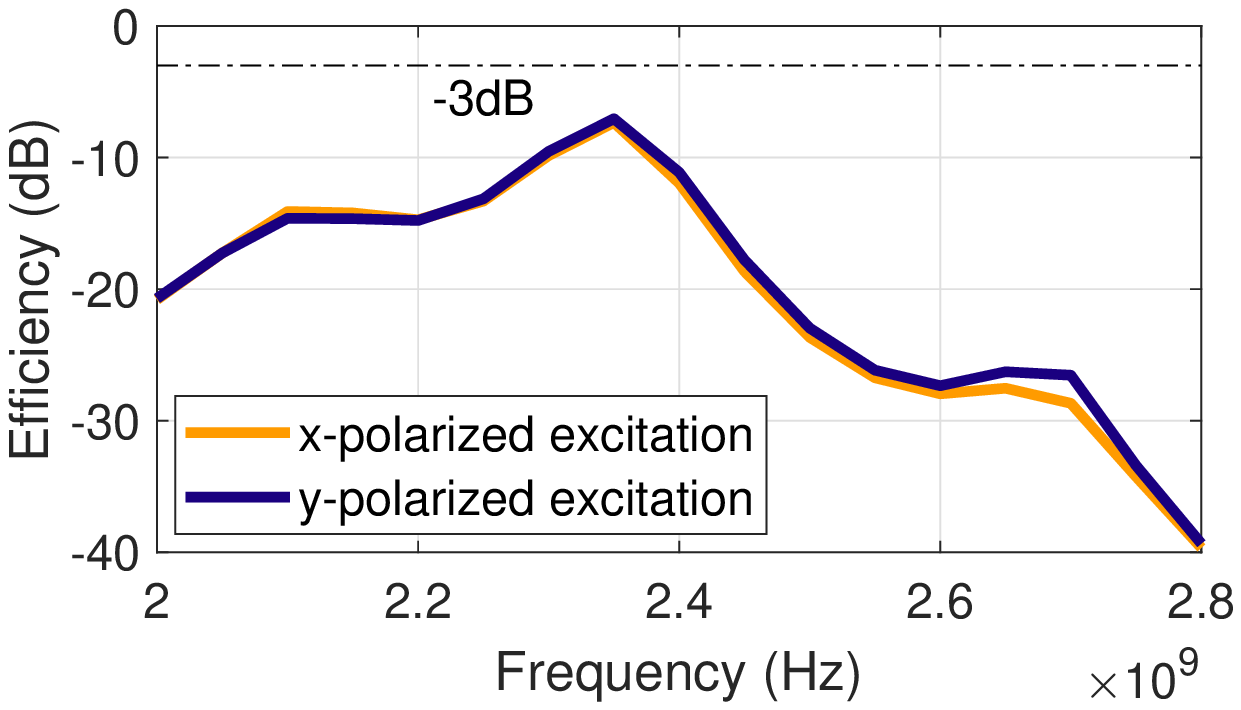}
  \caption{$S_{21}$ efficiency of cascaded polarization rotator layers using FR4 substrate~(loss tangent is $0.02$).}
  \label{FR4_efficiency}
  \end{minipage}
  \hspace{0.25cm}
  \begin{minipage}[c]{0.31\textwidth}
  \includegraphics[width=1\textwidth]{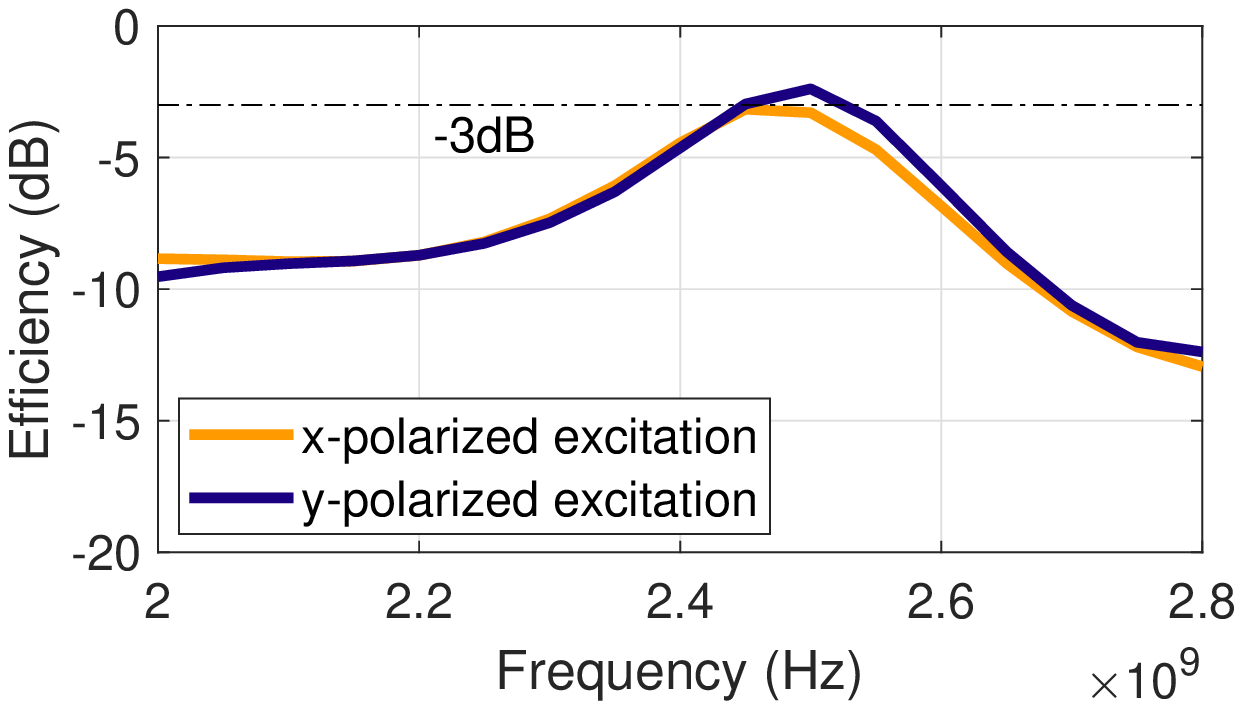}
  \caption{$S_{21}$ efficiency of \textbf{optimized} cascaded polarization rotator layers using FR4 substrate~(loss tangent is $0.02$). }
  \label{optimized_FR4_efficiency}
  \end{minipage}
\end{figure*}

% The key challenge in scaling a metasurface structure is determining the correct microstrip feature geometry for inductive and capacitive elements such that the scaled structure will be tuned for the desired frequency band. 
We can potentially obtain high transmission efficiency~(see Figure~\ref{Rogers_efficiency}) by directly scaling the circuit geometry of an existing $10$~GHz design~\cite{wu2019tunable} to $2.4$~GHz, but a key limitation is its use of an expensive, low-loss dielectric substrate (Rogers 5880)~\cite{wu2019tunable}. While this material achieves high transmission efficiency, it is cost prohibitive at scale. Instead, we choose a commodity FR4 substrate and characterize the behavior of the FR4 structure using an HFSS simulation environment to analyze critical parameters in the $2.4$~GHz ISM band.. The key problem is that FR4~($0.02$ dielectric loss tangent) causes much larger signal loss than Rogers 5880~($0.0009$ dielectric loss tangent), and thus severely decreases the transmission efficiency, as shown in Figure~\ref{FR4_efficiency}.

To reduce transmission loss, we simplify the structure of the tunable phase shifter layers, and decrease the thickness of FR4 by replacing it with an air gap since the dielectric loss tangent of air is 0. By comparing Figure~\ref{optimized_FR4_efficiency} and Figure~\ref{Rogers_efficiency}, we can see that our optimized structure made of cheap FR4 can achieve comparable transmission efficiency to more complex structures and expensive materials. We use fewer~(\emph{i.e.}, two) phase shifting layers made with thinner substrate; since the supported bandwidth of a phase shifter changes approximately linearly with the transmission line length, that is, the thickness of the substrate. Suppose the thickness of substrate is $\lambda/m$, the bandwidth can be represented as below~\cite{rao2015microwave}:
\begin{equation}\label{bandwidth}
\Delta f=f_{0}(2-\frac{m}{\pi}arccos[\frac{\Gamma}{\sqrt{1-\Gamma^{2}}}\frac{2\sqrt{Z_{I}Z_{L}}}{|Z_{L}-Z_{I}|}]),
\end{equation}
where $f_0$ is the design center frequency of phase shifter, $\Gamma$ is the maximum tolerable reflection coefficient, $Z_{I}$ and $Z_{L}$ are input impedance and load impedance, respectively. Our design achieves~($150$~MHz of bandwidth with efficiency $>$ $-5$~dB), which is wider than the target ISM frequency band that has less than $100$~MHz of bandwidth.

\begin{figure}[!t]
  \centering
  \vspace{-0.1cm}
  \includegraphics[width=0.47\textwidth]{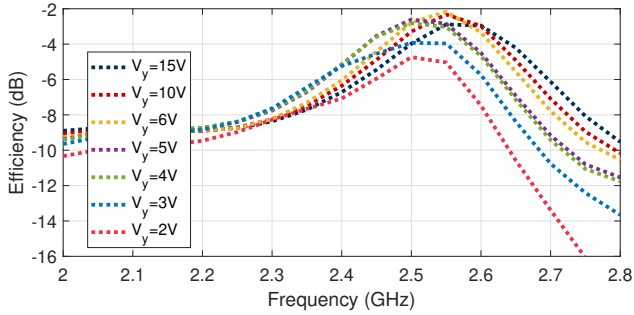}
  \caption{$S_{21}$ efficiencies under different voltage combinations of $X$ and $Y$ axes. The results show polarization can be controlled by changing the biasing voltages of phase shifter.}
  \label{polarization_results}
\end{figure}

\noindent\textbf{Estimating Polarization Efficiency.} 
 %Lumped capacitances ranging from $0.84$~pF to $2.41$~pF were used to approximate a varactor diode (SMV1233) used as part of an LC tank circuit for the $X$ and $Y$ planes; reverse bias voltages from $2$~V to $15$~V would realize these capacitance values. 
 A voltage controlled capacitance is used to actuate the tuning of the X and Y planes --- here we use an x-polarized incident wave as an example to show the polarization rotation results. The transmission efficiencies of the simulated frequencies under various voltage combinations are shown in Figure~\ref{polarization_results}, which are always higher than $-8$~dB in the $2.4-2.5$~GHz ISM frequency band.
The other set of measurements looks at how the polarization angle can be controlled by adjusting the lumped tuning capacitance used for the $X$ and $Y$ axis biasing layers. Varying this capacitance from $0.84$~pF to $2.41$~pF for both the $X$ and $Y$ axes resulted in a polarization rotation angle that varied between %$-37.8^{\circ}$ and $+60^{\circ}$ for x-polarized wave~(see Table~\ref{tab:x-rotator-degrees}), and for y-polarized wave, the rotation angle can be varied from 
$1.9^{\circ}$ and $48.7^{\circ}$~(see Table~\ref{tab:y-rotator-degrees}). We have also simulated the polarization rotator structure in the $900$~MHz band used for RFID and found comparable performance after additional scaling.

% \begin{table}[!t]
%   \caption{Simulated rotation degrees ($\theta_{r}$) for x-polarized wave.}
%   \label{tab:x-rotator-degrees}
%   \footnotesize
%   \begin{tabular}{l|c|ccccccc}
%     \hline
% 		\multicolumn{2}{c|}{\multirow{2}{*}{$\theta_{r}$ ($^{\circ}$)}} & \multicolumn{7}{c}{$V_{x} (V)$}  \\
% 		\cline{3-9}
% 		\multicolumn{2}{l|}{}
% 		& 2 & 3 & 4 & 5 & 6 & 10 & 15\\
% 		\hline
% 		%\hline
% 		\multirow{7}{*}{$V_{y} (V)$} & 2 & 5.6 & 23.0 & 38.1 & 44.8 & 49.9 & 60.0& 55.5  \\
%      	& 3 & 6.5 & 9.8 & 26.7 & 34.3 & 38.1 & 41.4 & 42.0 \\
% 		& 4 & 22.6 & 8.7 & 12.0 & 18.9 & 23.0 & 27.4 & 27.9  \\
% 		& 5 & 25.3 & 11.3 & 9.0 & 16.0 & 20.5 & 25.2 & 25.7  \\
% 		& 6 & 32.1 & 19.7 & 4.3 & 2.4 & 7.7 & 13.6 & 14.0  \\
% 		& 10 & 36.6 & 24.4 & 9.4 & 3.5 & 3.2 & 9.2 & 9.6  \\
% 		& 15 & 37.8 & 26.4 & 12.8 & 7.4 & 2.2 & 4.7 & 5.1  \\
% 	\hline
% \end{tabular}
% \end{table}

\begin{table}[!t]
  \caption{Simulated rotation degrees ($\theta_{r}$). }
  \label{tab:y-rotator-degrees}
  \footnotesize
  \begin{tabular}{l|c|ccccccc}
    \hline
		\multicolumn{2}{c|}{\multirow{2}{*}{$\theta_{r}$ ($^{\circ}$)}} & \multicolumn{7}{c}{$V_{x} (V)$}  \\
		\cline{3-9}
		\multicolumn{2}{l|}{}
	    & 2 & 3 & 4 & 5 & 6 & 10 & 15\\
		\hline
		%\hline
		\multirow{7}{*}{$V_{y} (V)$} & 2 & 11.6 & 26.1 & 36.8 & 41.0 & 44.3 & 48.3 & 48.7  \\
		& 3 & 6.5 & 12.4 & 26.6 & 32.2 & 35.2 & 38.6 & 39.2 \\
		& 4 & 23.0 & 4.9 & 10.9 & 17.3 & 20.8 & 25.0 & 25.6  \\
		& 5 & 27.0 & 9.3 & 7.4 & 14.0 & 18.0 & 22.6 & 23.2  \\
		& 6 & 41.8 & 25.0 & 7.9 & 2.1 & 4.2 & 10.2 & 10.7  \\
		& 10 & 45.8 & 30.0 & 13.7 & 7.9 & 2.8 & 5.1 & 5.6  \\
		& 15 & 48.2 & 33.1 & 18.2 & 12.9 & 7.3 & 1.9 & 2.0  \\
	\hline
\end{tabular}
\end{table}

\begin{algorithm}[!t]
\SetAlgoLined
    \caption{Biasing Voltage Sweep} \label{algorithm: voltage-control}
 \textbf{Input:} {Number of iterations: $N$; Number of voltage tuning steps for $X$ and $Y$ axes per iteration: $T$}  
     
\textbf{Initialization:} {Voltage sweep range of $X$ and $Y$ axes in first iteration~($n=1$): $[V_{x,1}^{min},V_{x,1}^{max}]=[0,30]$,
     $[V_{y,1}^{min},V_{y,1}^{max}]=[0,30]$ } 
     
\For {$n=1,..., N$} 
{\For {$\tau=1,..., T$}
 { $V_{x,n,\tau}=V_{x,n}^{min}+(\tau-1)(V_{x,n}^{max}-V_{x,n}^{min})/T$ \\
$V_{y,n,\tau}=V_{y,n}^{min}+(\tau-1)(V_{y,n}^{max}-V_{y,n}^{min})/T$ }
 
\If {Received signal power at voltage combination $(v_{x,n,\tau}, v_{y,n,\tau})$ is strongest}
 { $V_{x,n+1}^{min}=V_{x,n,\tau}-(V_{x,n}^{max}-V_{x,n}^{min})/T$ \\
     $V_{x,n+1}^{max}=V_{x,n,\tau}$ \\
     $V_{y,n+1}^{min}=V_{y,n,\tau}-(V_{y,n}^{max}-V_{y,n}^{min})/T$ \\
     $V_{y,n+1}^{max}=V_{y,n,\tau}$}
%\EndIf
}
 \textbf{Output:} {Optimal voltage combination: $(V_{x,N,\tau}, V_{y,N,\tau})$}
     
\end{algorithm}

\begin{figure*}[htbp]
  \centering
  \subfigure[Signal power as a function of Tx rotation without metasurface.]{
  \includegraphics[width=0.23\textwidth]{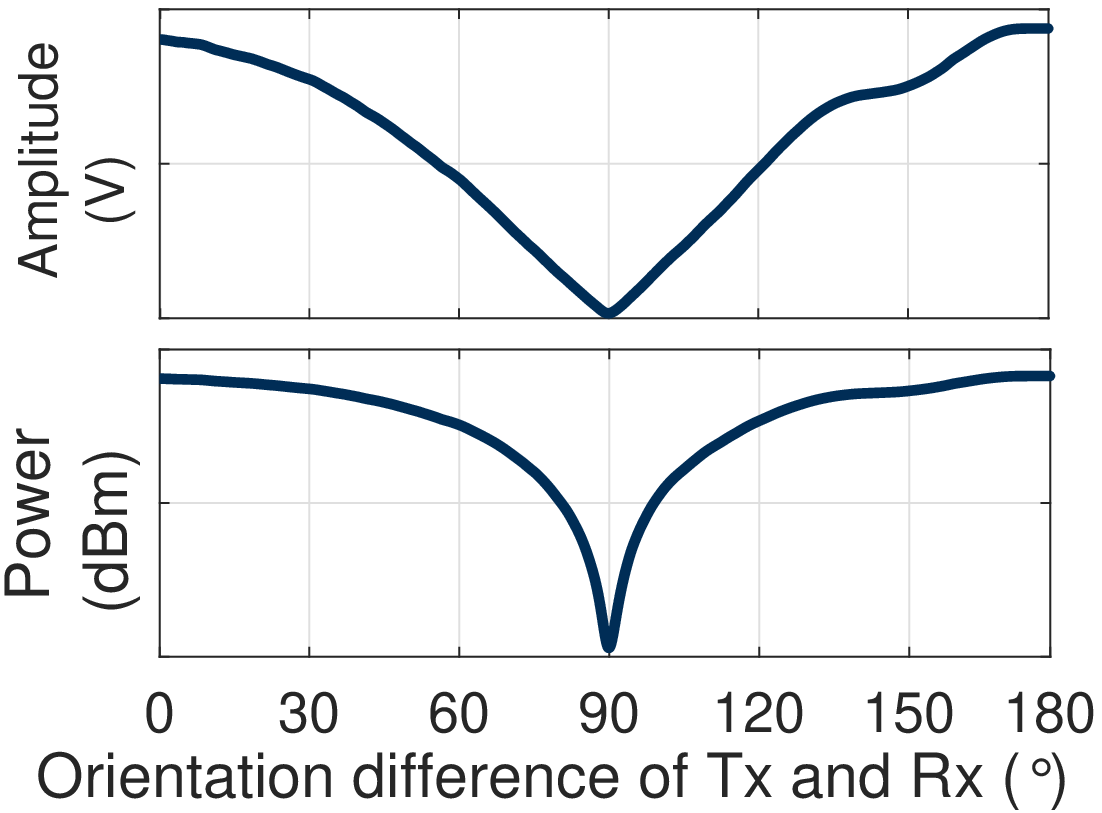}}
  \hspace{0.1cm}
  \subfigure[Signal power over voltage combination of metasurface.]{
  \includegraphics[width=0.23\textwidth]{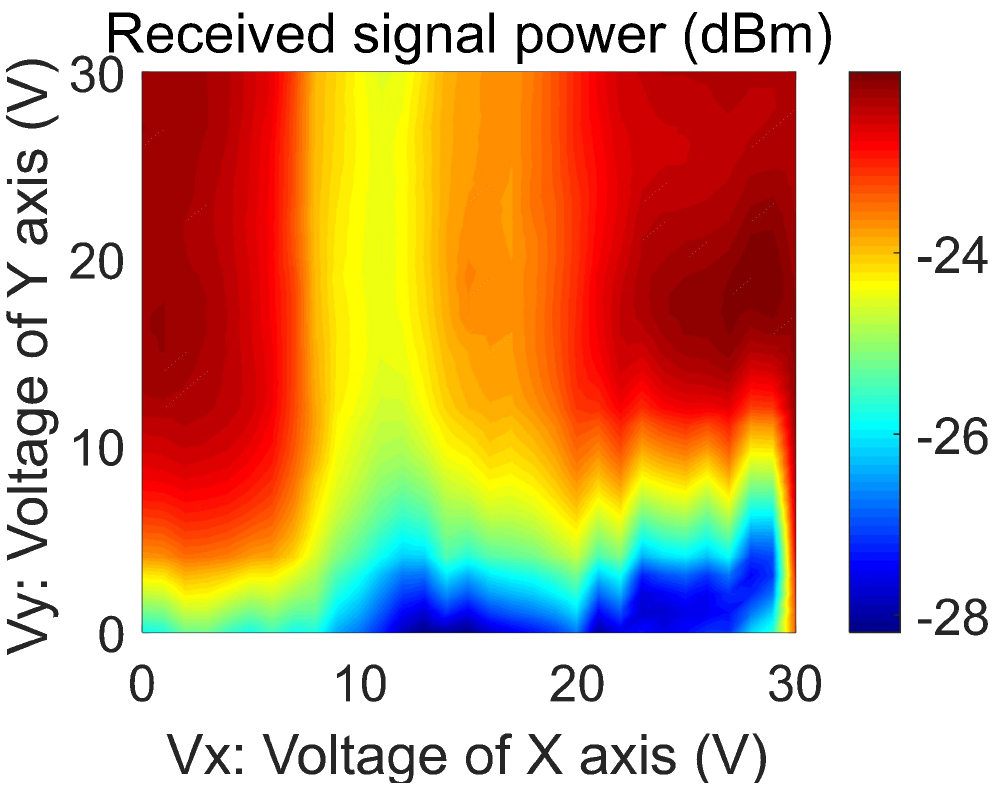}}
  \hspace{0.1cm}
  \subfigure[Minimum and maximum polarization rotation angles computation.]{
  \includegraphics[width=0.23\textwidth]{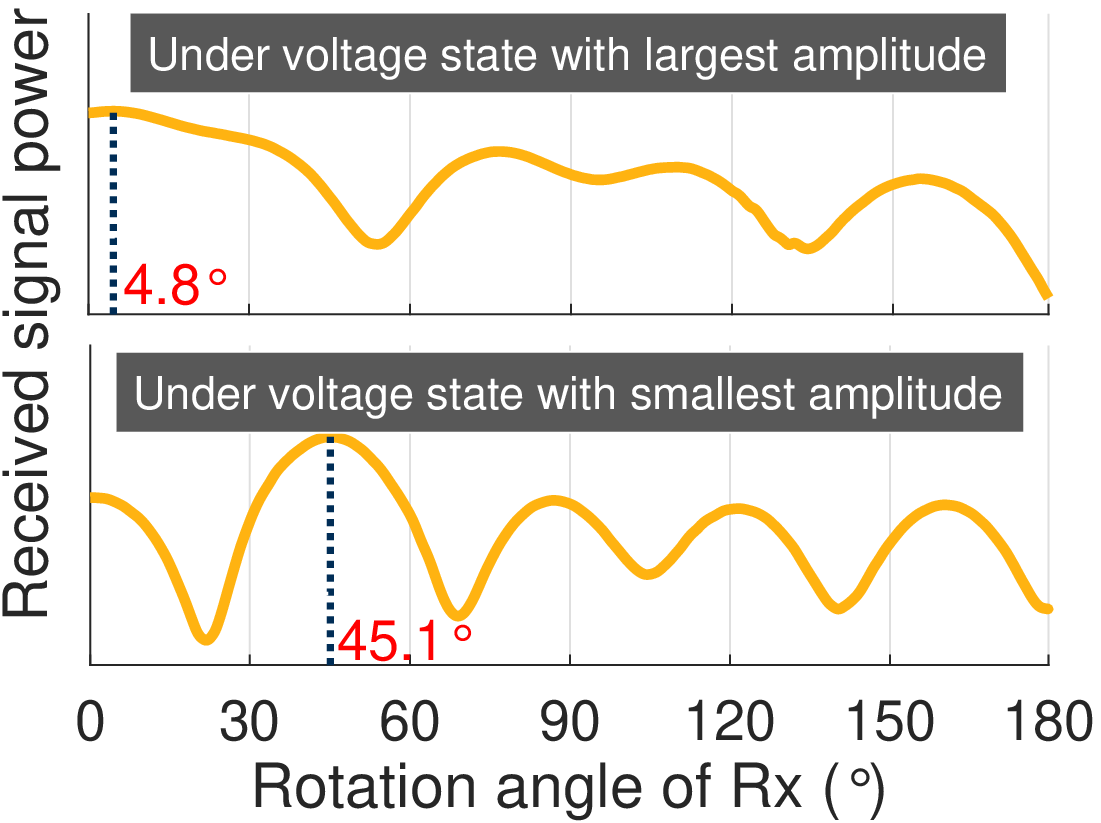}}
  \hspace{0.1cm}
  \subfigure[Estimated polarization rotation angle over voltage combination.]{
  \includegraphics[width=0.23\textwidth]{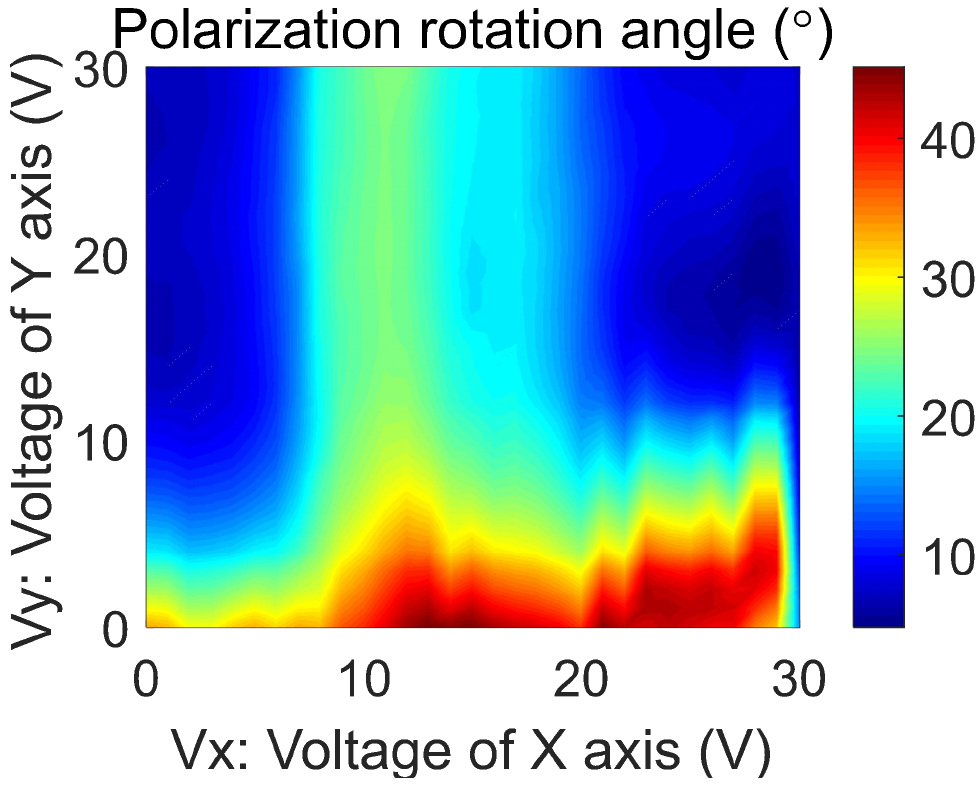}}
  %\vspace{-0.2cm}
  \caption{Polarization rotation degree estimation according to the received signal power. %Figure~(a) shows the signal power change pattern over orientation difference between Tx and Rx without metasurface. 
 % The rotation degree estimates in Figure~(d) are obtained according to the power change slope in (a), power measurements in (b), and minimum and maximum rotation angle estimates in (c). 
 The acceptable ``band'' around the optimal value shown in (c) is chosen by considering multipath interference.} 
  \label{rotation-angle}
\end{figure*}

\subsection{Metasurface Control}\label{section:bias-voltage-control}
To enable polarization rotation control, we need to change the capacitance of the $X$ and $Y$ axis phase shifters; in our design this is accomplished by changing the bias voltage of the integrated varactor diodes (SMV1233) in the $X$ and $Y$ polarities, which in turn changes their capacitance and thus phase. All diodes in a given polarization are controlled using the same bias voltage; we use a programmable power supply for this purpose and these voltages can be as high as $30$~V to account for errors induced during fabrication and assembly, hence we set $0-30$~V as the voltage sweep range of the $X$ and $Y$ axes. The power supply is connected to a desktop computer through a USB interface, and is controlled by a Python script that uses the Virtual Instrument Software Architecture (VISA) standard with a maximum voltage switching frequency of $50$~Hz. With a voltage step of $1$~V, the full scan across both X and Y axes takes $\sim$~$30$ seconds, which prevents real-time applications.

% \begin{algorithm}[t]
% \SetAlgoLined
%     \caption{Biasing Voltage Sweep} \label{algorithm: voltage-control}
%  \textbf{Input:} {Number of iterations: $N$; Number of voltage tuning steps for $X$ and $Y$ axes per iteration: $T$}  
     
% \textbf{Initialization:} {Voltage sweep range of $X$ and $Y$ axes in first iteration~($n=1$): $V_{rx,1}=[v_{x,1}^{min},v_{x,1}^{max}]$,
%      $V_{ry,1}=[v_{y,1}^{min},v_{y,1}^{max}]$ } 
     
% \For {$n=1,..., N$} 
%  {Voltage sweep for $X$ and $Y$ axes in range of $V_{rx,n}$ and $V_{ry,n}$, with step of $V_{s,n}=(v_{x,n}^{max}-v_{x,n}^{min})/T$ }
 
% \If {Received signal power at voltage combination $(v_{x,n,t}, v_{y,n,t})$ is strongest}
%  { $n=n+1$, return $V_{rx,n}=[v_{x,n,t}-V_{s,n},v_{x,n,t}]$ and
%      $V_{ry,n}=[v_{y,n,t}-V_{s,n},v_{y,n,t}]$}
% %\EndIf

%  \textbf{Output:} {Optimal voltage combination: $(v_{x,N,t}, v_{y,N,t})$}
     
% \end{algorithm}

To reduce the sweep time, we start with a coarse-grained voltage sweep then increase the resolution of control as summarized in Algorithm~\ref{algorithm: voltage-control}. Specifically, we define $N$ as the number of iterations, and $T$ as the number of voltage adjustments per iteration. %$[V_{x,n}^{min},V_{x,n}^{max}]$ and $[V_{y,n}^{min},V_{y,n}^{max}]$ are the voltage sweep ranges of the $X$ and $Y$ axis in the $n^{\text{th}}$ iteration.
The time cost in $n^{\text{th}}$ iteration is $0.02\times N\times T^{2}$. We empirically set $T$ to $5$ and $N$ to $2$ , according to the switching speed and the voltage resolution of the programmable power supply. After $N$ iterations across the $X$ and $Y$ axes, we can determine the optimal voltage combination to yield the strongest received signal.

% \noindent\textbf{Synchronization between rotator and receiver.} \blue{For real-time polarization correction and communication link optimization, it is necessary to correlate the currently received sample with the bias voltage state, so that we can determine an optimal voltage combination that enables strongest received signal power.} Instead of involving another dedicated device~\cite{fan2018energy}, we achieve this by exploiting the fact that the sampling rate of receiver and the voltage switch speed of power supply are both constant over time. So we can label the received sample $X_{t}$ at time $t$ with voltage state $\{V_{x,t}, V_{y,t}\}$ as below:
% \begin{equation}
% \{V_{x,t}, V_{y,t}\}=\{V_{x,0}+\frac{VD_{x}}{T_{s}}\times (t-t_{d}), V_{y,0}+\frac{VD_{y}}{T_{s}}\times (t-t_{d})\}, 
% \end{equation}
% where $V_{x,0}$ and $V_{y,0}$ represent the voltages of $X$ and $Y$ axis at initial time $t=0$, $VD_{x}$ and $VD_{y}$ are the voltage difference between two adjacent voltage values of $X$ and $Y$ axes, $T_{s}$ is the time cost of a single voltage change, $t_{d}$ is the start time difference between receiver and power supply.

\noindent\textbf{Synchronization between rotator and receiver.} For real-time polarization correction and communication link optimization, it is necessary to correlate the currently received sample with the bias voltage state, so that we can determine an optimal voltage combination that enables the strongest received signal power. In our prototype, for simplicity we directly connect the receive antenna to the voltage supply, which allows us to assume the voltage switch speed and receiver sampling rate have a constant relationship over time. This allows us to relate received signal samples with the voltage state of \systemname at a given time instance. A full implementation can have the receiver explicitly send channel state information to the controller, as in previous work~\cite{li2019towards, dunna2020scattermimo}.

 One important aspect of the design we want to highlight is that the leaking current of our metasurface is as low as $15$~nA, which means the metasurface does not need a large battery or significant power from the AC mains to keep it powered; it can maintain operation with a modestly sized capacitor.

\begin{figure*}[!t]
  \centering
  %\vspace{-0.1cm}
  \subfigure[Prototype.]{
  \includegraphics[width=0.15\textwidth]{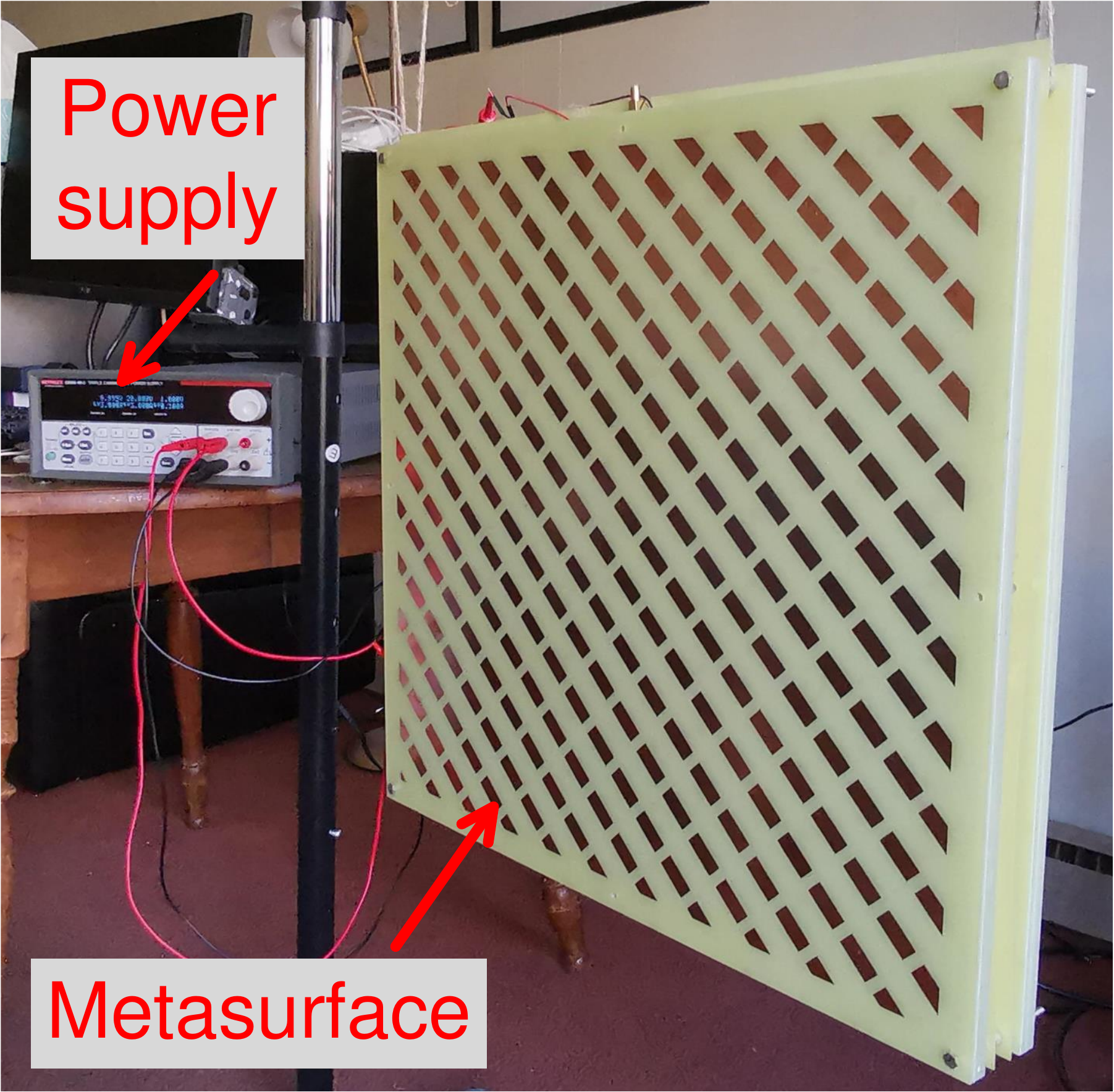}}\hspace{0.03cm}
  \subfigure[QWP outer layer.]{
  \includegraphics[width=0.15\textwidth]{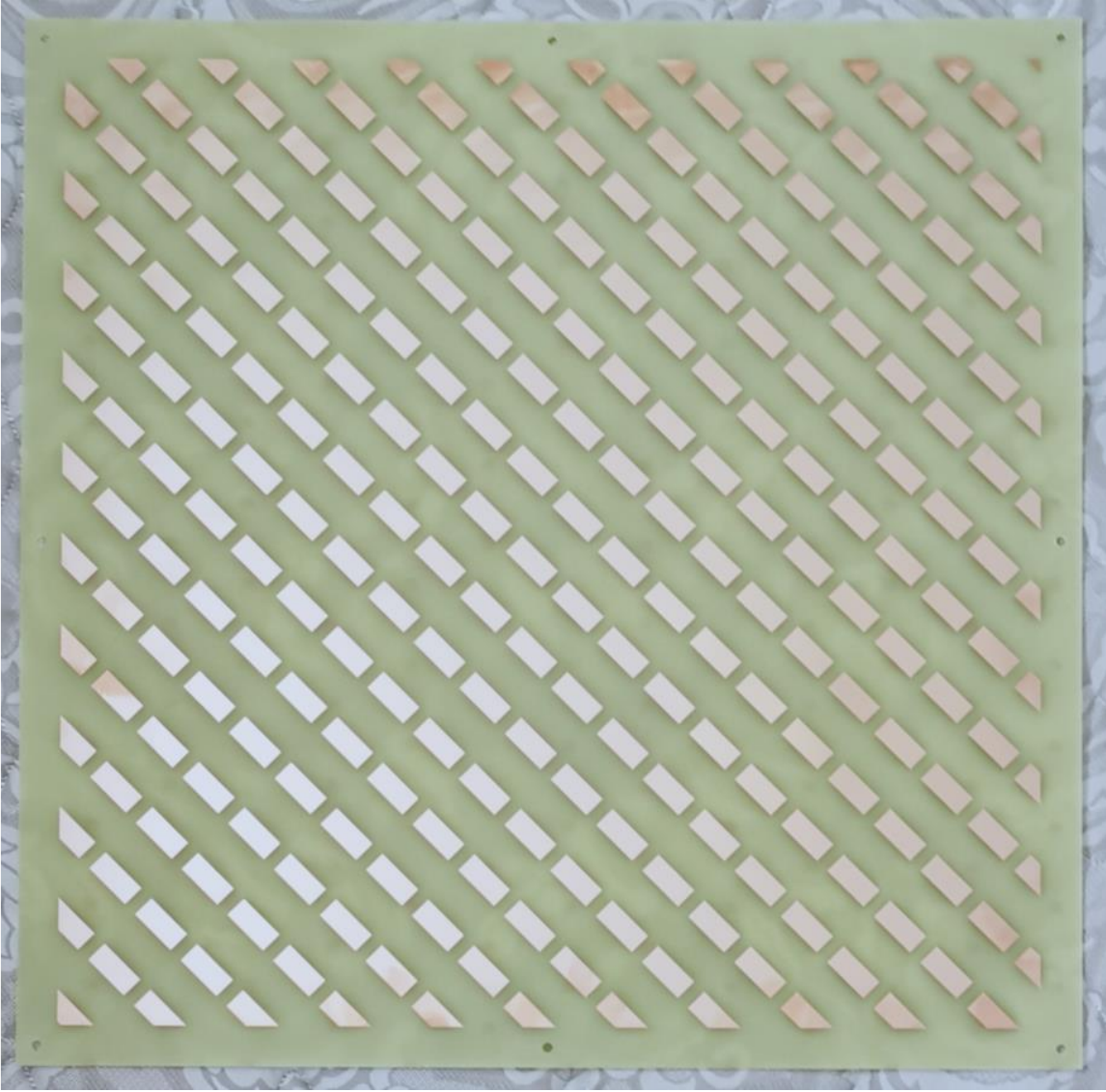}}
  \hspace{0.03cm}
  \subfigure[QWP inner layer.]{
  \includegraphics[width=0.15\textwidth]{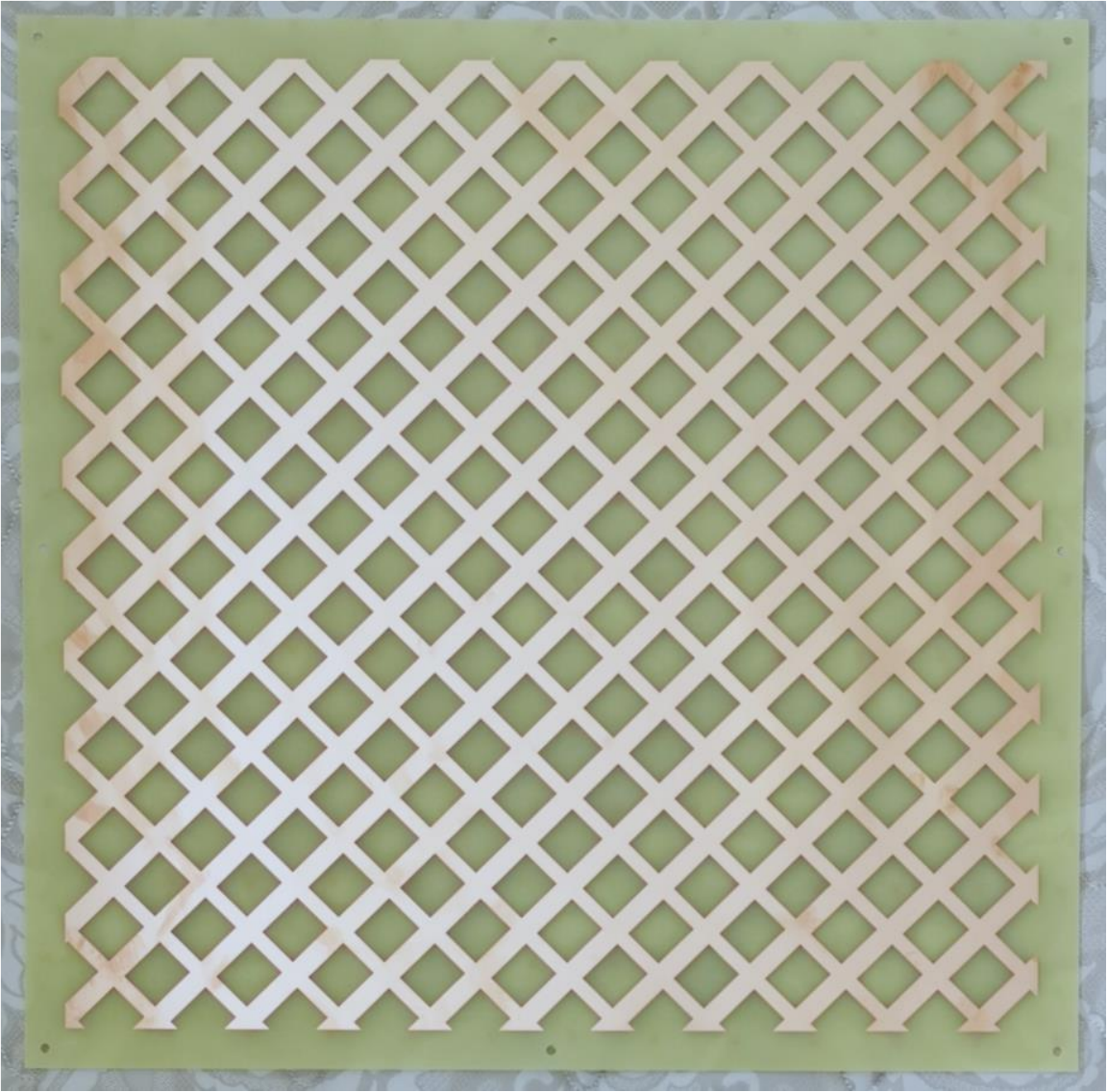}}
  \hspace{0.03cm}
  \subfigure[BFS layer.]{
  \includegraphics[width=0.15\textwidth]{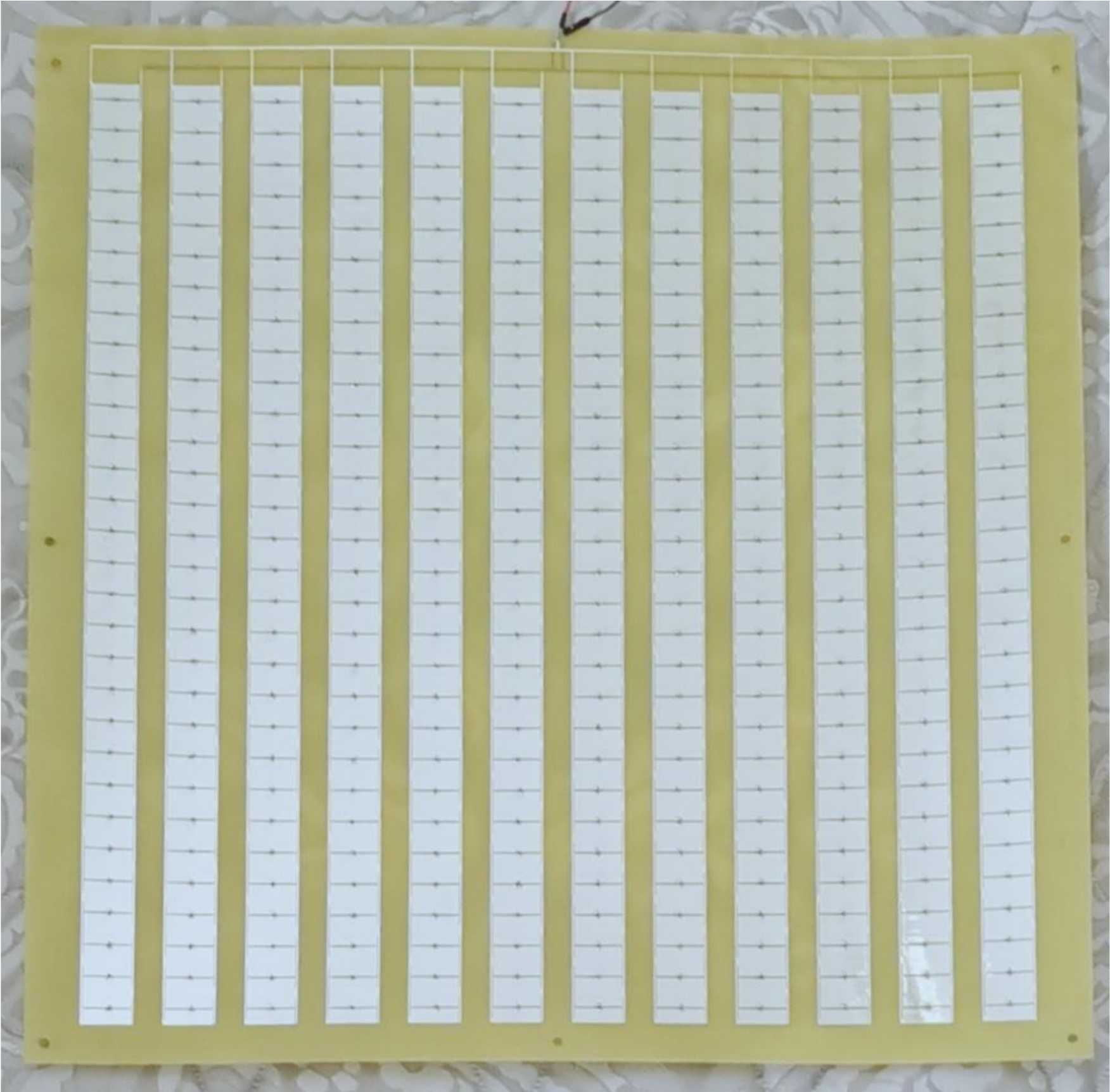}}
  \hspace{0.03cm}
  \subfigure[Evaluation setups.]{
  \includegraphics[width=0.33\textwidth]{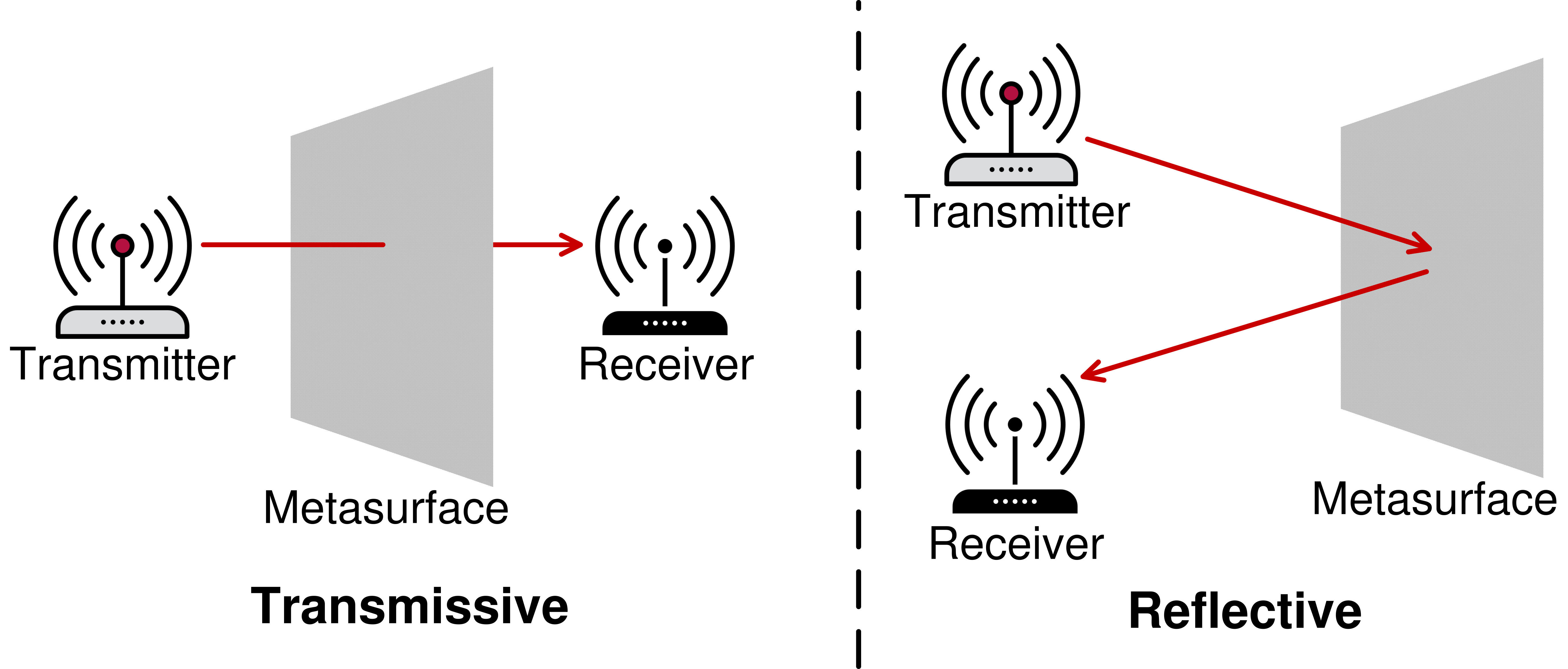}}
  %\vspace{-0.2cm}
  \caption{\systemname prototype and evaluation setups. (a-d) show the PCBs with diodes embedded in the BFS layer. (e) presents the two experimental setups used in the evaluation. The first is a transmissive configuration where endpoints are placed on either side of \systemname; the second is a reflective configuration where both endpoints are placed on the same side of \systemname.
  }
  \label{prototype}
\end{figure*}

\subsection{Polarization Rotation Degree Estimation}\label{section:mapping}
Besides increasing SNR, we can also sense the relative orientation of the two endpoints with \systemname. Here we present our approach for rotation angle estimation according to the received signal power reported by the endpoint receiver, no matter the transmitter-receiver distance; as distances between the transmitter and receiver become comparable to the size of the metasurface (See Figure~\ref{overview}), a fraction of the signal can bypass the metasurface without polarization rotation, and this results in less overall perceived rotation at the receiver. According to our benchmark experimental result plotted in Figure~\ref{rotation-angle}~(a), we observed that the received signal power~(before the dBm conversion) can be approximated as a linear change with the orientation difference between transmitter and receiver. When we perform measurements across a full voltage sweep, we can get the maximum potential improvement to signal power. To obtain the polarization rotation angle for an unknown transmitter-receiver distance~(\emph{i.e.}, the potential power improvement over an orientation sweep is unknown), the key is determining the minimum and maximum polarization rotation angles. We take the following steps.

\noindent\textbf{Step 1:} Fix the receiver at the same orientation with the transmitter, by rotating the receiver to find an orientation $\theta_{0}$ where the received power is largest.

\noindent\textbf{Step 2:} Sweep across voltage combinations $V_{min}$ and $V_{max}$ corresponding to min and max powers, respectively (\emph{i.e.}, parallel and orthogonal polarizations).

\noindent\textbf{Step 3:} Set the voltage state to the two searched combinations, respectively. At each voltage state, rotate the receiver by $180^{\circ}$ to find the new orientation where the power is strongest. The two new orientations of $V_{min}$ and $V_{max}$ can be defined as $\theta_{max}$ and $\theta_{min}$ as shown in Figure~\ref{rotation-angle}~(c). The differences of the receiver's initial orientation and two new orientations $|\theta_{0}-\theta_{min}|$ and  $|\theta_{0}-\theta_{max}|$ correspond to the minimum and maximum polarization rotation angles, respectively. 

The antenna that needs to be rotated is fixed on a turntable and rotated via remote control~\cite{Turntable}. From the experimental results of the match setup shown in Figure~\ref{rotation-angle}~(b-d), we can see that the polarization rotation angle varies between $5^{\circ}-45^\circ$ during the voltage sweep.

\section{Implementation and Experimental Setup}
\noindent\textbf{Metasurface.} We fabricated the metasurface with a total surface area of $48\times48~cm^{2}$ and a thickness of $5$~cm, including $180$ functional units~(Figure~\ref{prototype}~(a-d)). The biasing voltages of the metasurface are provided by a programmable power supply~(TektronixSeries 2230G~\cite{tektronix}) through two DC channels, as shown in Figure~\ref{prototype}~(a). %We connect the power supply to a Linux PC through a USB interface and control the power supply with a Python script that uses the Virtual Instrument Software Architecture (VISA) standard.
\systemname utilizes $720$ varactor diodes (SMV1233), costing $\sim$~$\$0.50$~each. The total cost of \systemname for all PCB layers is  $\sim\$540$, resulting in a total cost of $\sim\$900$. Given economies of scale, the unit cost can be reduced to $\$2$ when there are more than $3000$ units per PCB.

\noindent\textbf{Experimental setup.} For controlled experiments, we utilize one USRP N210 software-defined radio with a UBX-40 daughterboard as the ISM signal transceiver, operating at a default center frequency of $2.44$~GHz. The transmitter and receiver antennas are separated by a specified distance. We experiment with both directional~\cite{Alfa} and omni-directional antennas~\cite{Highfine}.
We configure and control the USRP using the GNU radio software development
toolkit~\cite{gnuradio} run on a PC. The transmitter continuously transmits a cosine waveform over $500$~KHz, while the sampling rate of the receiver is $1$~MHz. We also evaluate \systemname using low-cost Wi-Fi and Bluetooth devices~(the same setup as benchmark experiments as shown in Figure~\ref{motivation}). Additionally, we perform an experiment with a pair of GIGABYTE mini-PCs~\cite{BRIX} with Intel 5300 wireless cards, 
to evaluate performance over a larger frequency range~(\emph{i.e.}, $20$~MHz, including $52$ OFDM subcarriers). %The communication is between a Wi-Fi AP~\cite{router} and a cheap ESP 8266-based Arduino board~\cite{esp8266}. The AP can send data at a rate up to $340$ Mbps.

We perform both through-surface (\textit{transmissive}) and surface-reflection (\textit{reflective}) experiments as shown in Figure~\ref{prototype}~(e).
In transmissive experiments, the metasurface is placed between the transmitter and receiver. In reflective experiments, the transmitter and receiver are placed on the same side of the metasurface.
In each experiment, the baseline received signal power without the metasurface is measured by averaging $30$ seconds of received samples, and the maximum signal power with the metasurface is obtained after a fast sweep of voltages as detailed in \S~\ref{section:bias-voltage-control}. 
To avoid multipath effects confounding the performance behavior of \systemname, we cover the test area with RF absorbing material, and use directional antennas by default in USRP-based experiments.

\section{Evaluation}
In this section, we conduct extensive experiments to evaluate the performance of \systemname. We first answer how metasurface improves the transmissive signal power in polarization mismatch setup. Then we analyze the relationship between signal enhancement induced by the metasurface across a number of parameters including transmitted power, multipath effect, antenna directionality and operating frequency. We also evaluate \systemname's performance for practical low-cost IoT communication links. In addition, we validate \systemname's ability to enhance a reflected signal, and demonstrate the influence of the proposed metasurface structure for sensing. 

\noindent\textbf{Performance metrics.} 
We measure signal strength at the receiver as our performance metric, since this directly characterizes the benefit of polarization rotation. An increase in the received power usually translates to a throughput improvement. While it is common to measure link throughput directly, the size limit of our current prototype makes it challenging to characterize link throughput in diverse settings. %We leave this to future work once we scale to a large deployment. While we believe the throughput would improve according to to increased receive power since this is one of the most important factors.

\begin{figure*}[!t]
  \centering
  %\vspace{-0.1cm}
  \subfigure[24cm Tx-Rx distance.]{
  \includegraphics[width=0.23\textwidth]{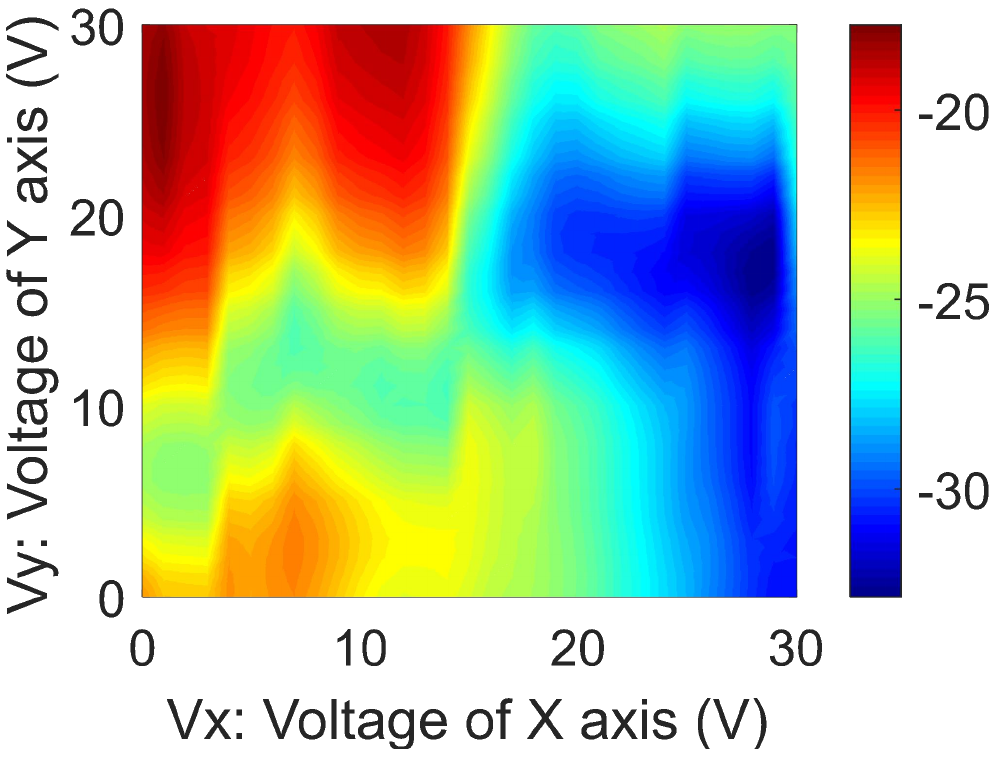}}
  \hspace{0.1cm}
  \subfigure[30cm Tx-Rx distance.]{
  \includegraphics[width=0.23\textwidth]{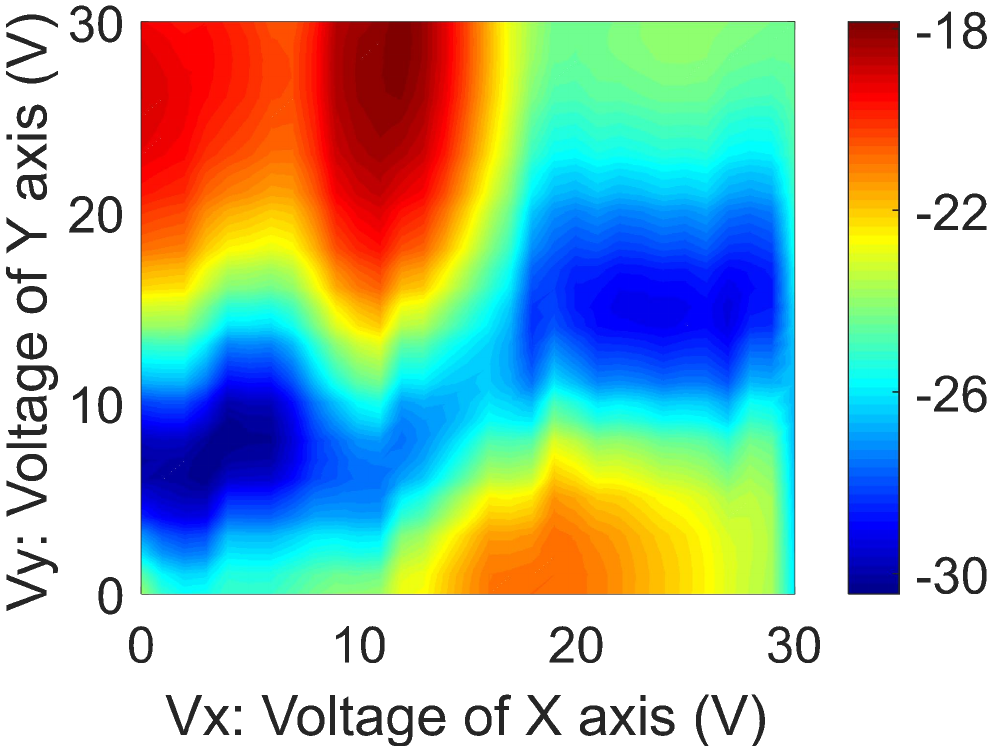}}
  \hspace{0.1cm}
  \subfigure[36cm Tx-Rx distance.]{
  \includegraphics[width=0.23\textwidth]{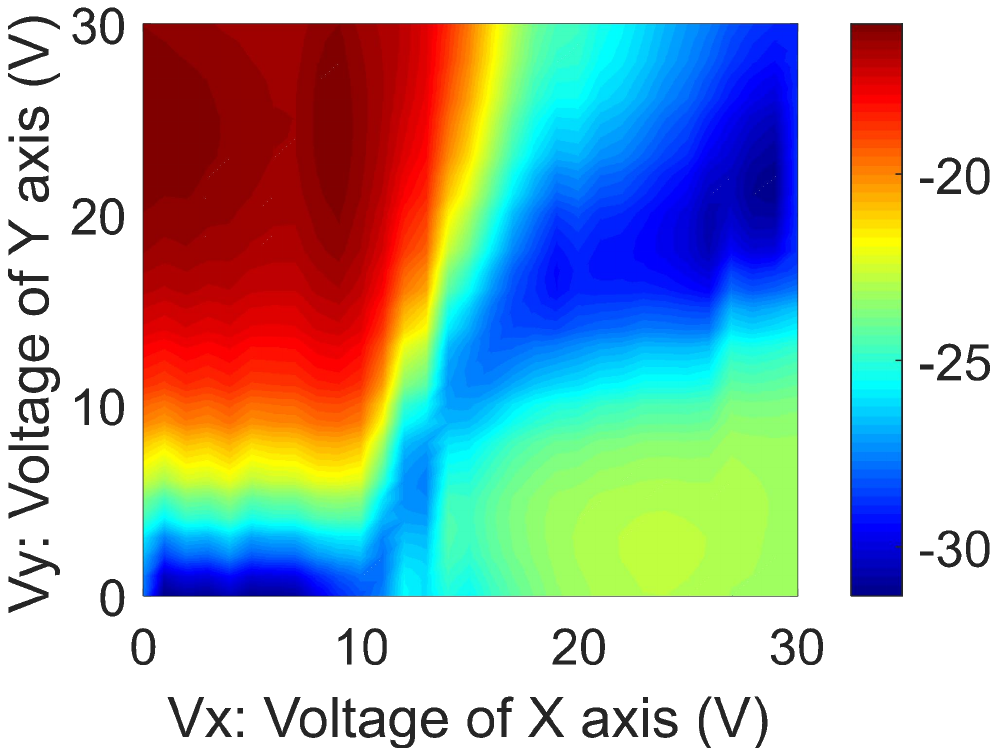}}
  \hspace{0.1cm}
  \subfigure[42cm Tx-Rx distance.]{
  \includegraphics[width=0.23\textwidth]{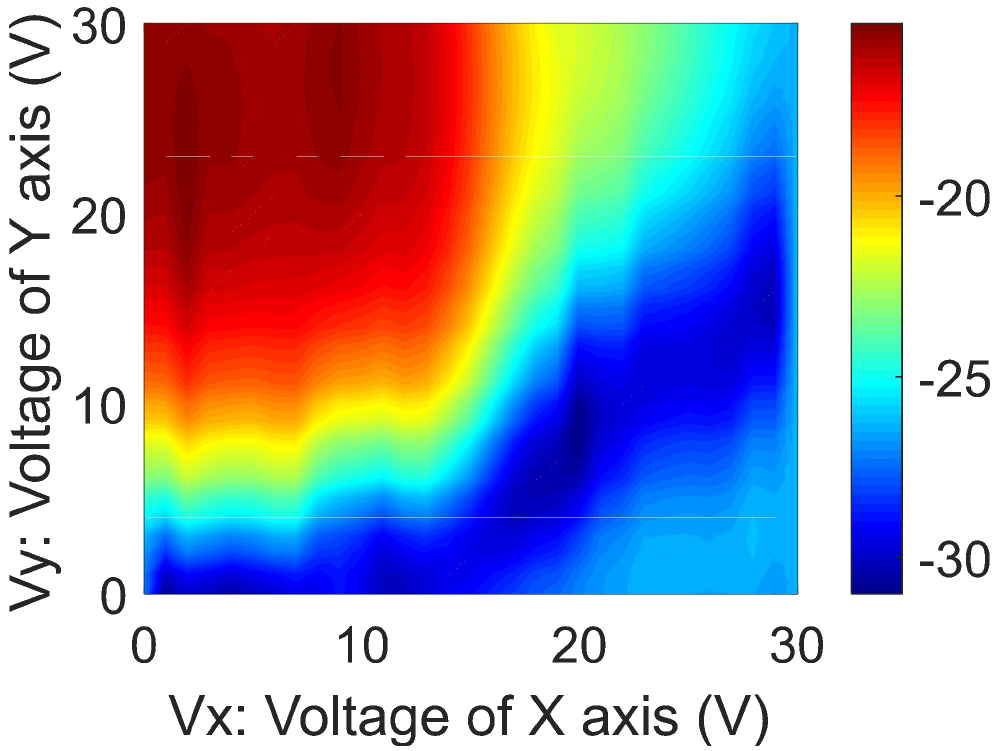}}
  \hspace{0.1cm}
  \subfigure[48cm Tx-Rx distance.]{
  \includegraphics[width=0.23\textwidth]{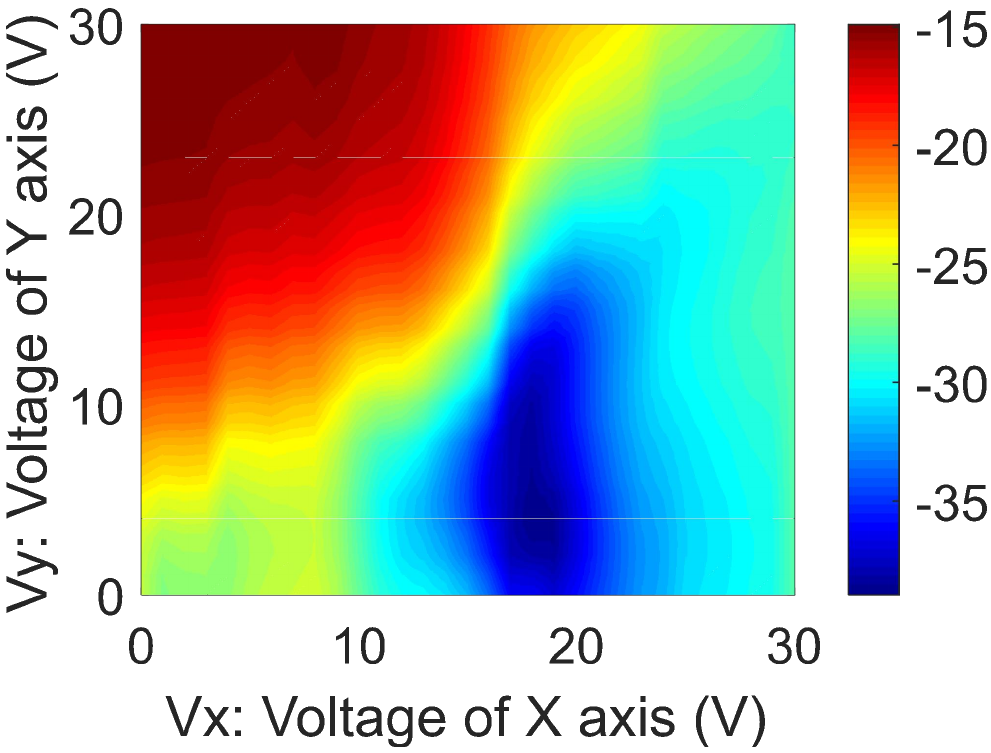}}
  \hspace{0.1cm}
  \subfigure[54cm Tx-Rx distance.]{
  \includegraphics[width=0.23\textwidth]{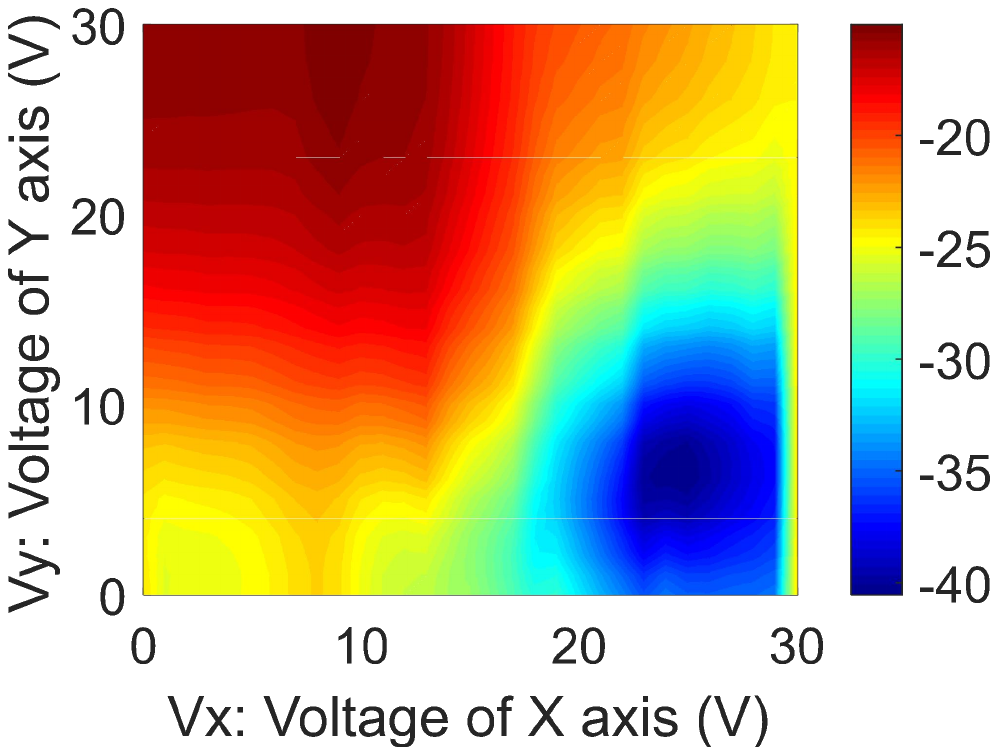}}
  \hspace{0.1cm}
  \subfigure[60cm Tx-Rx distance.]{
  \includegraphics[width=0.23\textwidth]{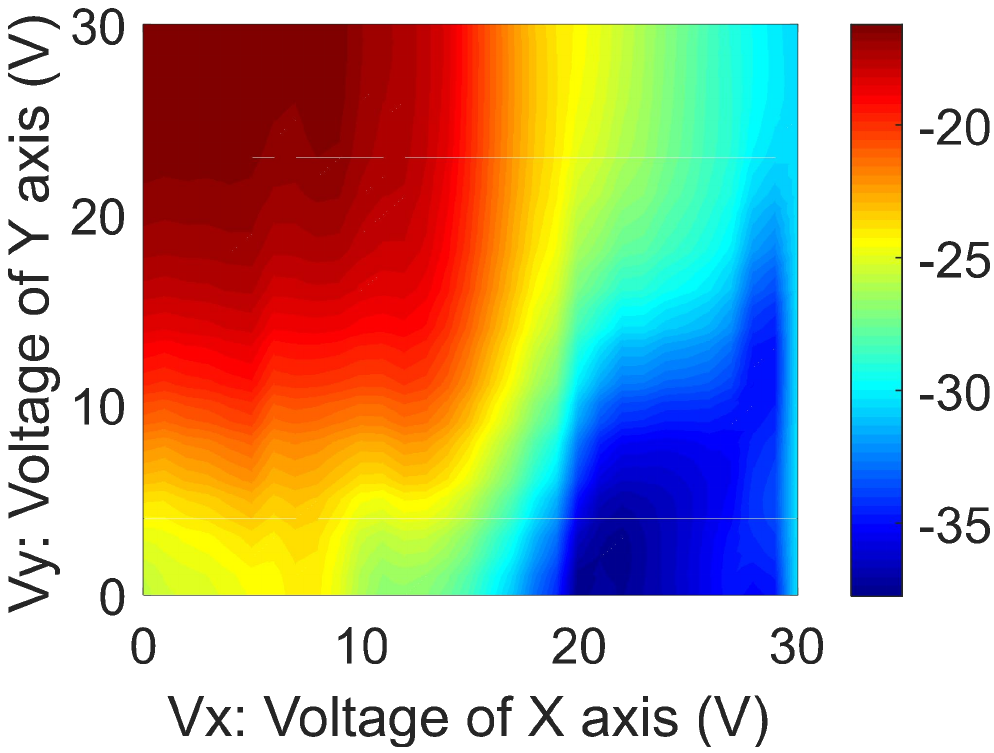}}
  \hspace{0.1cm}
 \subfigure[Polarization rotation degree.]{
  \includegraphics[width=0.23\textwidth]{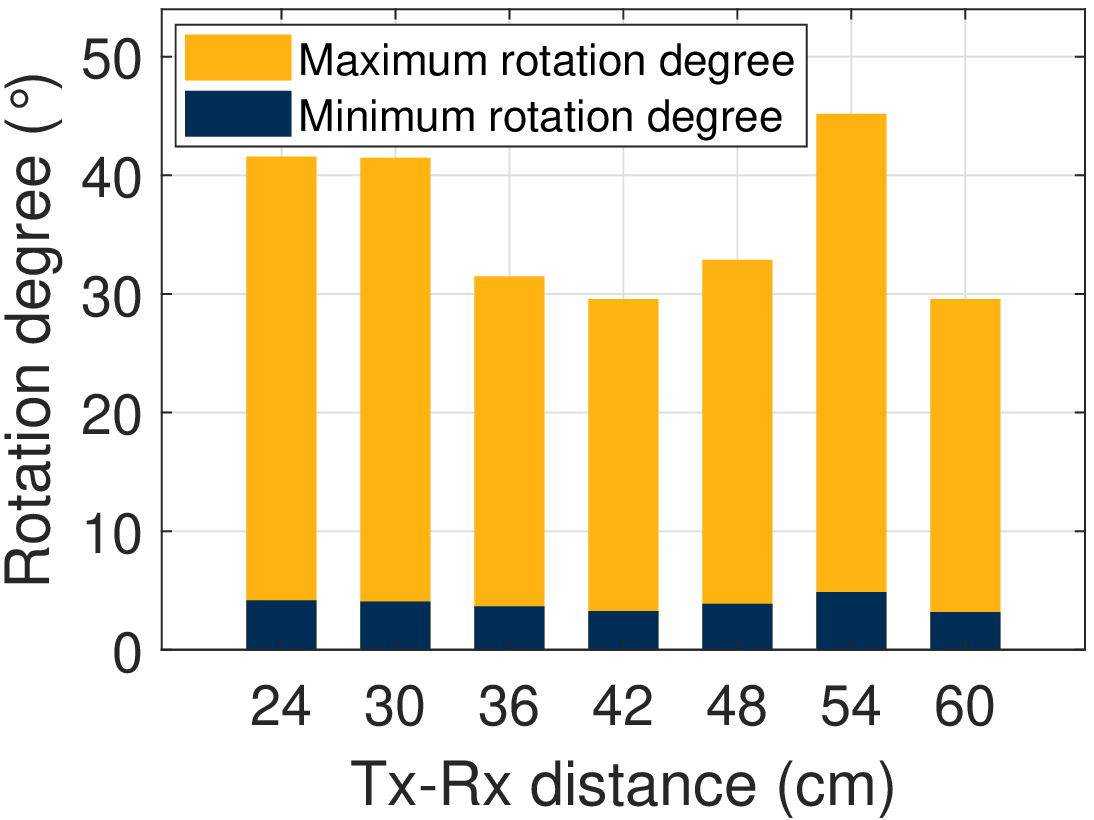}}
  %\vspace{-0.2cm}
  \caption{Measurements with metasurface under polarization mismatch setup. (a-g) show the received signal power heatmap with different voltage combinations. (h) presents the maximum polarization rotation degree caused by metasurface.} 
  \label{fig:transmissive-heatmaps}
\end{figure*}

\begin{figure*}[!t]
   \centering
   %\vspace{-0.1cm}
  \begin{minipage}[c]{0.47\textwidth}
  \includegraphics[width=1\textwidth]{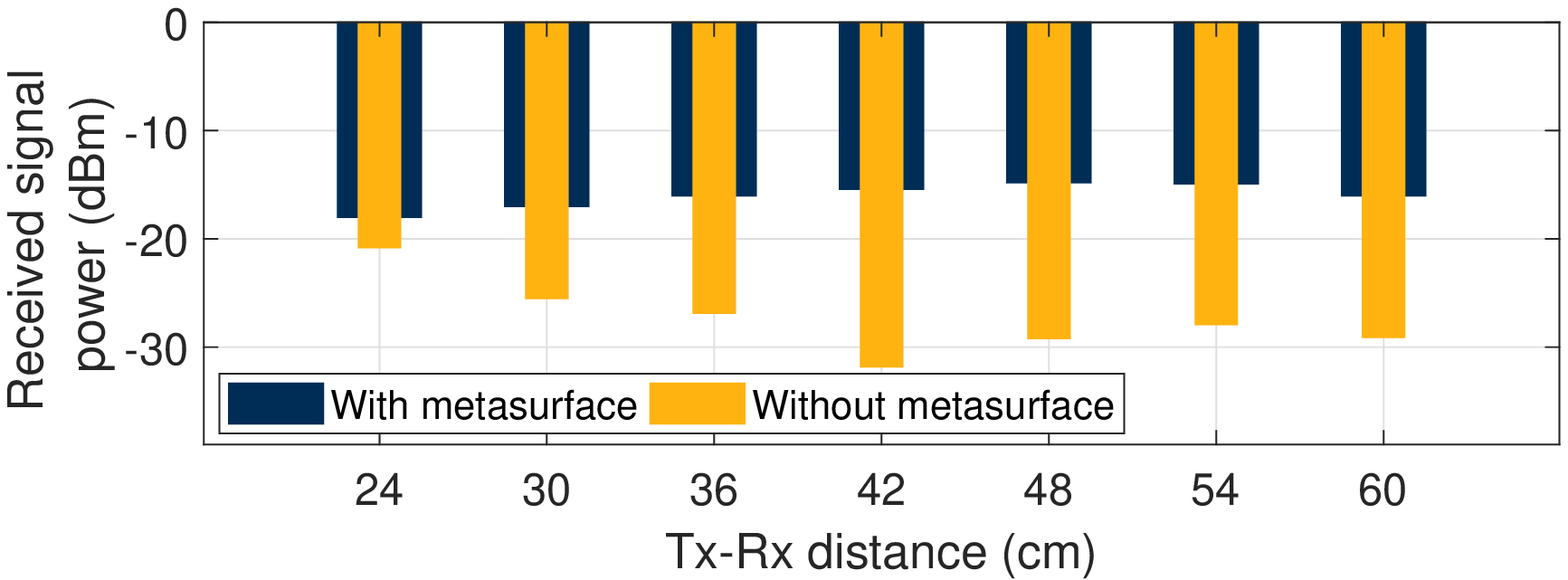}
  \caption{Received signal power with/without metasurface in polarization mismatch setup. }
  \label{transmissive-improvement}
  \end{minipage}
  \hspace{0.6cm}
  \begin{minipage}[c]{0.47\textwidth}
  \includegraphics[width=1\textwidth]{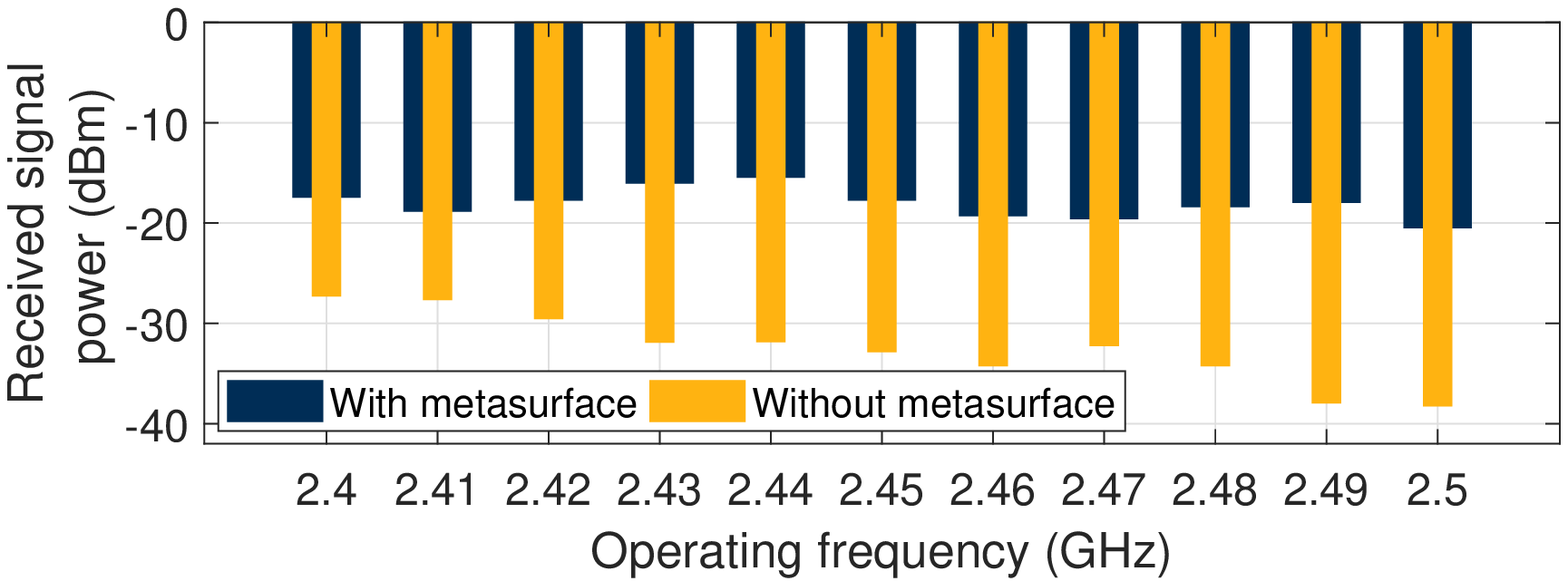}
  \caption{Power improvement VS. operating frequency in polarization mismatch setup.}
  \label{varying-frequency}
  \end{minipage}
\end{figure*}

\subsection{Transmissive Operation}
\subsubsection{Transmissive Signal Enhancement}
%In this transisissive experiment, we answer three questions: 1) How does the received signal power change according to the bias voltages of the metasurface? 2) What amount of polarization rotation can the metasurface achieve during its biasing voltage sweep? 3) How much can the signal power be increased by using the metasurface? 
To verify \systemname's ability to rotate the polarization of transmissive signals, we conduct experiments with the metasurface under different transmitter\hyp{}receiver (Tx-Rx) distances
(from $24$~cm to $60$~cm by half wavelength steps of $6$~cm). The transmitter and receiver are placed orthogonally such that they are in a mismatched polarization configuration. In each experiment, we measure the received signal power across a full sweep of voltage combinations (both $V_{x}$ and $V_{y}$ vary from $0-30$~V). Figure~\ref{fig:transmissive-heatmaps}~(a-g) show how the received signal power changes with different voltage combinations at each Tx-Rx distance and how the maximum achieved rotation angle diminishes as the distance becomes comparable to the size of the surface. The signal power changes significantly with changes in biasing voltage. We also find the mapping between these voltages as the rotation shifts gradually with respect to Tx-Rx distance. Figure~\ref{fig:transmissive-heatmaps}~(h) shows the polarization rotation degree measured by the proposed method in \S~\ref{section:mapping}. We find that the metasurface can rotate the polarization over a range of $3^{\circ}-45^{\circ}$, which allows the metasurface to correct for a significant amount of mismatch. To understand the signal improvements provided by the metasurface, we also measure the signal power in mismatch configuration with no metasurface present as a baseline. By comparing the results with and without the metasurface as depicted in Figure~\ref{transmissive-improvement}, we can see that the metasurface enhances the transmissive signal power by up to $15$~dB, which extends the potential transmission distance by up to $5.6\times$ according to the Friis equation~\cite{Friis}.

\subsubsection{Performance Benchmarks}
\noindent\textbf{Validating operational bandwidth.}
 In order to evaluate \systemname's performance over the entire ISM frequency band, we conduct experiments that vary the operating frequency from $2.4$~GHz to $2.5$~GHz by steps of $0.01$~GHz. We measure the maximum signal power with and without the metasurface. From the results shown in Figure~\ref{varying-frequency}, we can see that \systemname enables $>10$~dB signal enhancement across the entire ISM frequency band, when compensating for polarization mismatch (orthogonal antenna orientation). This indicates that \systemname has potential for optimizing IoT communication links with protocols including Wi-Fi, Bluetooth and Zigbee.

\begin{figure*}[!t]
   \centering
   %\vspace{-0.1cm}
   \begin{minipage}[c]{0.47\textwidth}
   \subfigure[Omni-directional antenna.]{
  \includegraphics[width=0.47\textwidth]{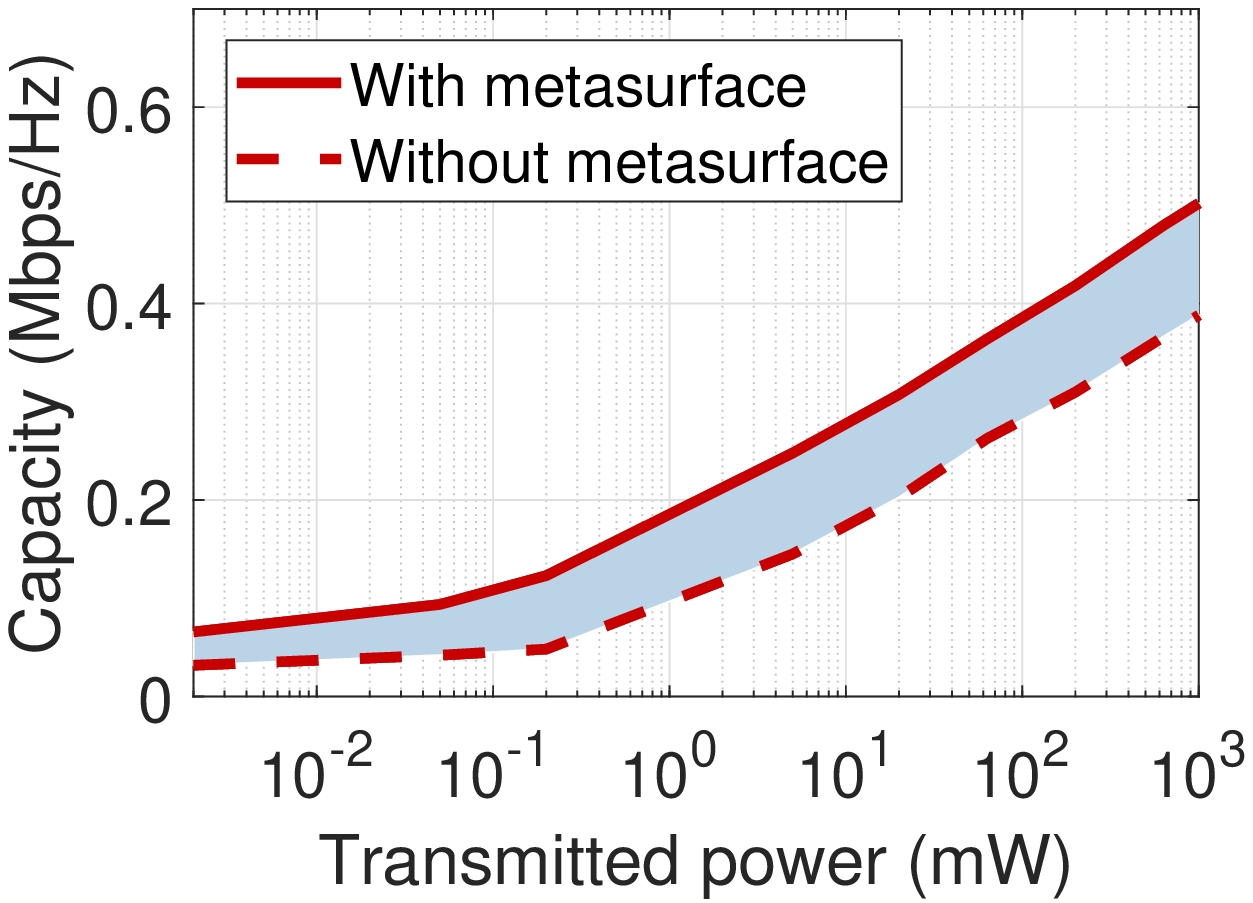}}
  \hspace{0.1cm}
  \subfigure[Directional antenna.]{
  \includegraphics[width=0.47\textwidth]{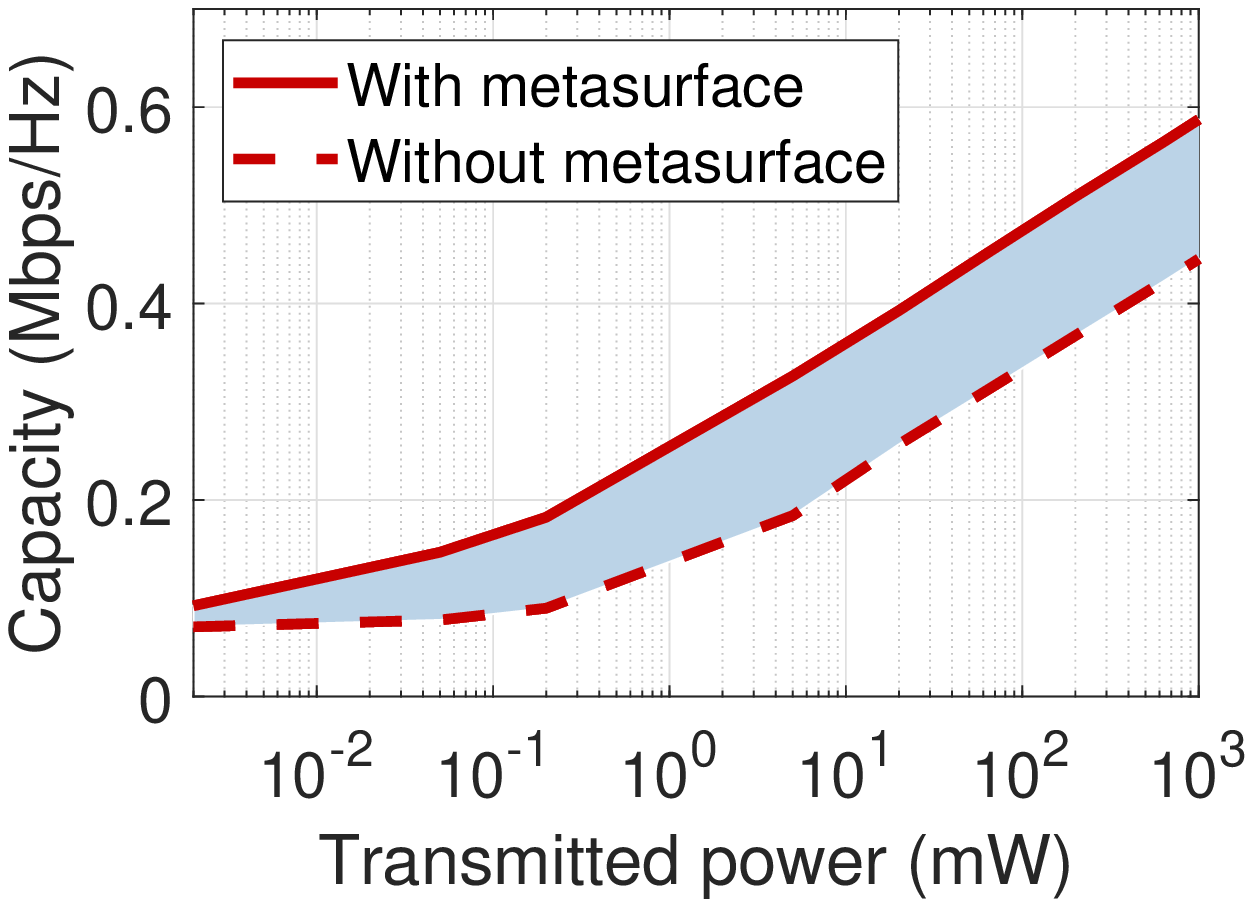}}
  %\vspace{-0.2cm}
  \caption{Channel capacity with varying incident power. We eliminate multipath by using RF absorbing material.}
  \label{with-absorber}
  \end{minipage}
  \hspace{0.5cm}
  \begin{minipage}[c]{0.47\textwidth}
   \subfigure[Omni-directional antenna.]{
  \includegraphics[width=0.47\textwidth]{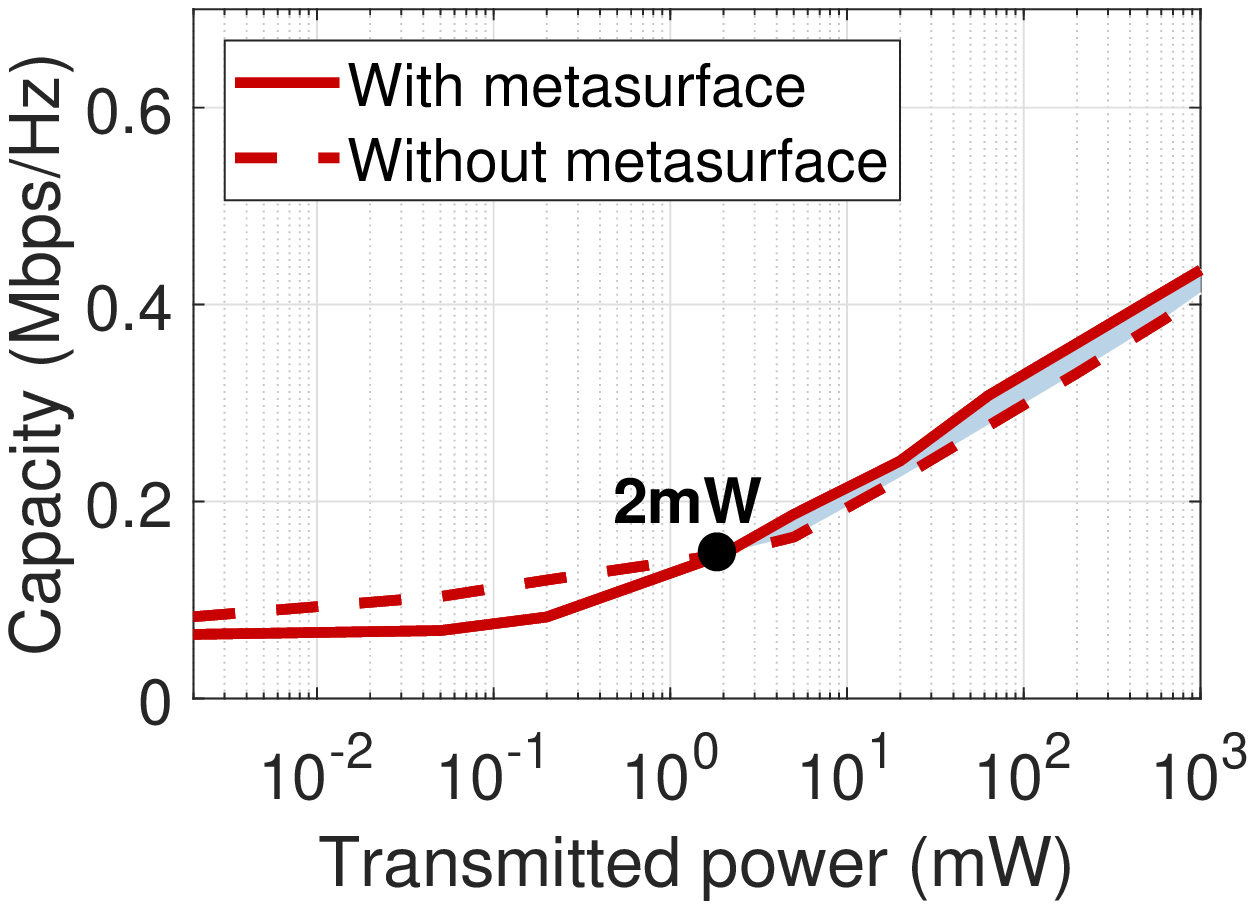}}
  \hspace{0.1cm}
  \subfigure[Directional antenna.]{
  \includegraphics[width=0.47\textwidth]{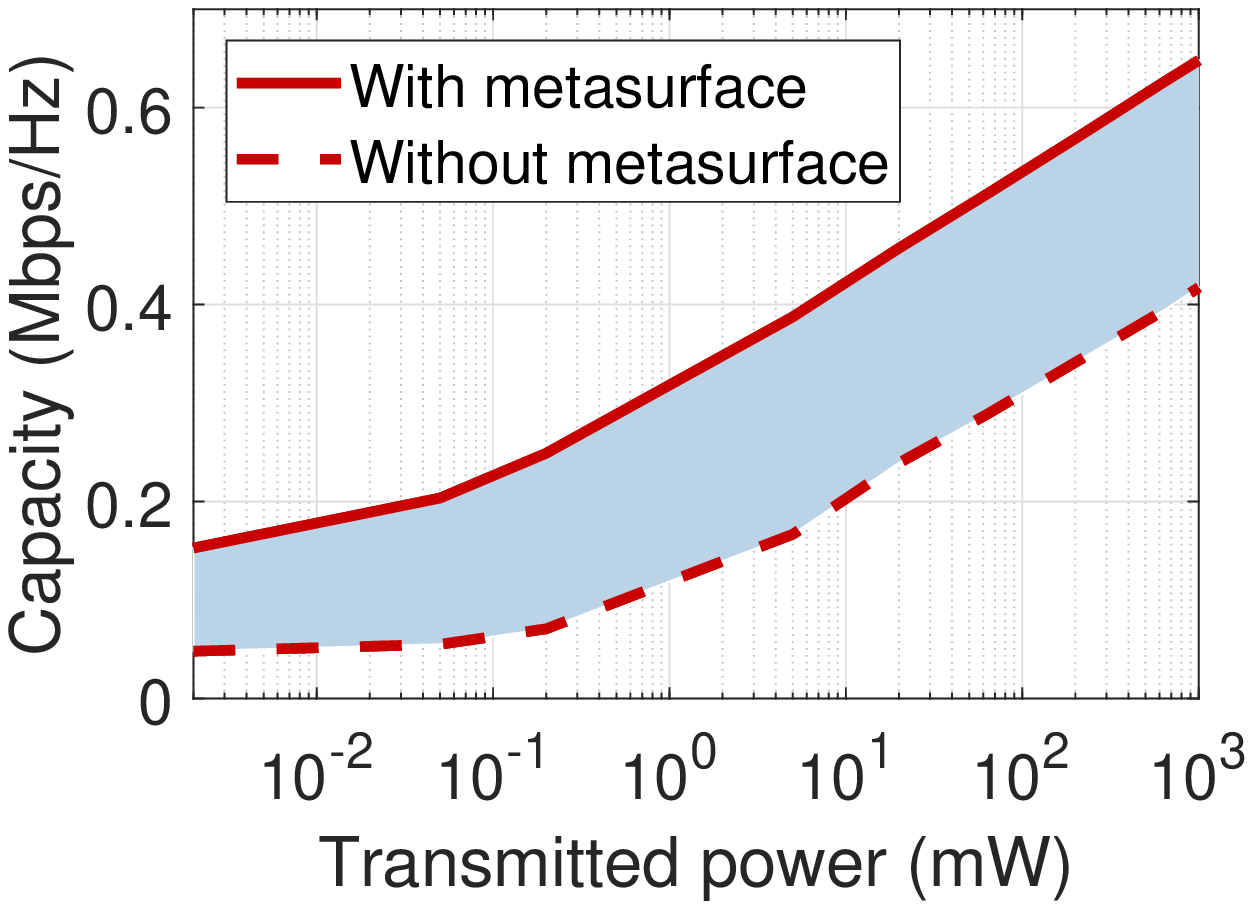}}
  %\vspace{-0.2cm}
  \caption{Experimental results in rich multipath environment~(laboratory) without using RF absorbing material. %(a) indicates that \systemname is able to improve signal power as long as the transmit power is higher than $2$~mW even with multipath effect.
  }
  \label{without-absorber}
  \end{minipage}
\end{figure*}

\begin{figure*}[!t]
   \centering
   \vspace{-0cm}
   \begin{minipage}[c]{0.47\textwidth}
   \subfigure[Wi-Fi.]{
  \includegraphics[width=0.47\textwidth]{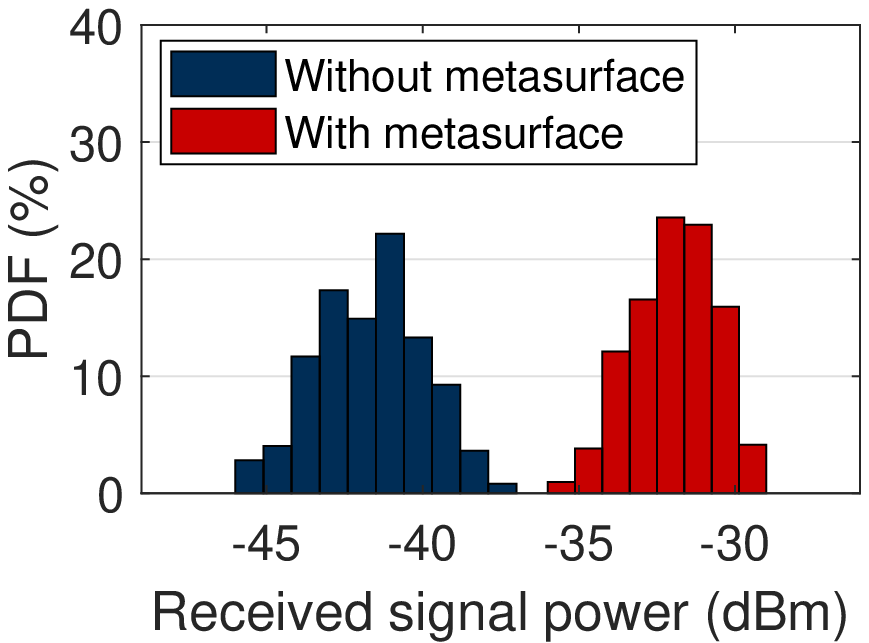}}
  \hspace{0.1cm}
  \subfigure[Bluetooth.]{
  \includegraphics[width=0.47\textwidth]{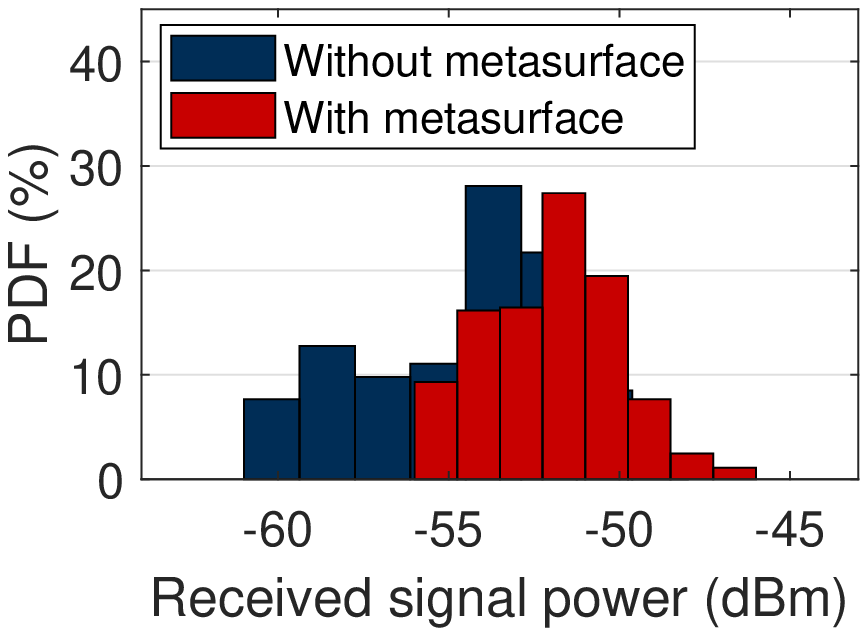}}
  \vspace{-0.3cm}
  \caption{Experimental results of low-cost IoT devices in polarization mismatch setup.}
  \label{IoT-devices}
  \end{minipage}
  \hspace{0.5cm}
  \begin{minipage}[c]{0.47\textwidth}
  \includegraphics[width=1\textwidth]{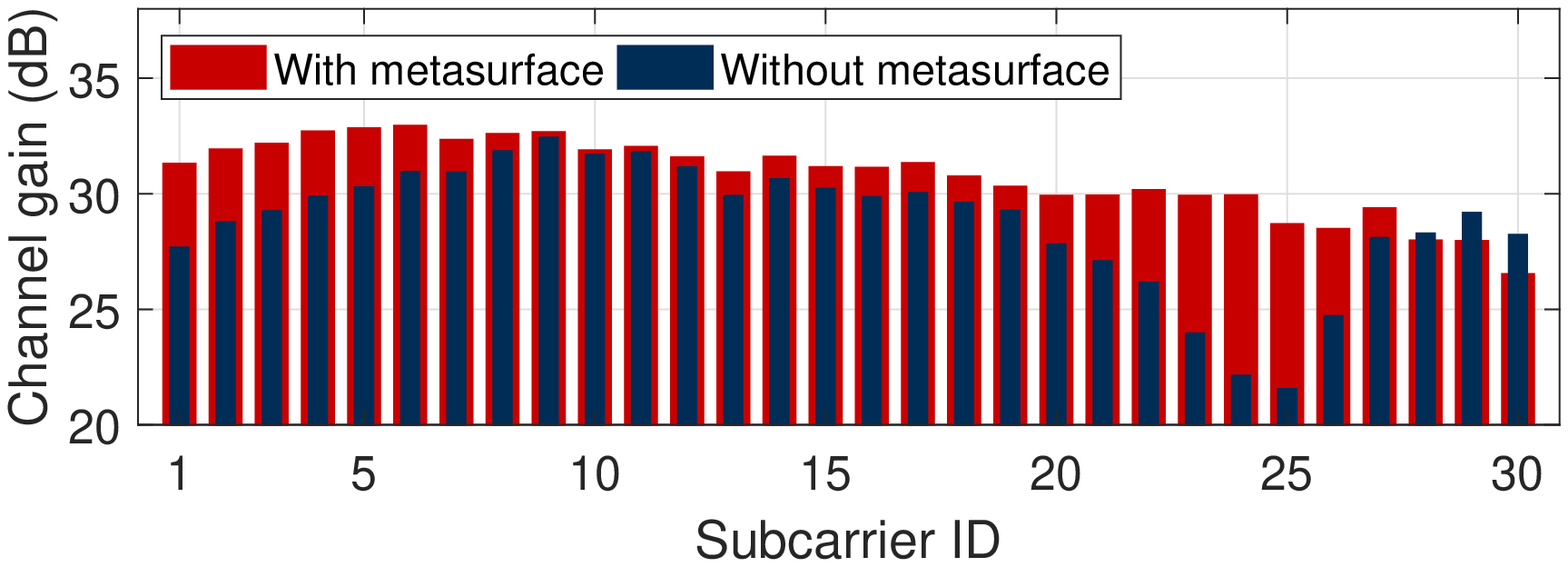}
  \caption{Experimental results of OFDM frequency bins in polarization mismatch setup.}
  \label{OFDM}
  \end{minipage}
\end{figure*}

\noindent\textbf{Impact of incident power on performance gains.}
In this experiment, we sweep through transmit power settings to understand how incident power affects the performance improvement (measured in terms of channel capacity enhancement) provided by the metasurface --- lower transmit powers could potentially be dominated by loss within the metasurface. The capacity is calculated according to the SNR measurement and channel bandwidth. We perform experiments with both directional~\cite{Alfa} and omnidirectional~\cite{Highfine} antennas on the  transceiver. Figure~\ref{with-absorber} shows that the capacity initially increases slowly with transmitting power for both the directional and omnidirectional antennas, due to the loss induced by the metasurface; the ``cut-off" transmit power for the metasurface to provide benefit is as low as $2$~mW.

\noindent\textbf{Impact of multipath.}
In these experiments we seek to understand the impact of multipath propagation on \systemname's performance. We perform experiments in an indoor lab environment without absorber material. We also measure the channel capacity by using two types of antennas at different transmit power. By comparing the results of Figure~\ref{without-absorber} with Figure~\ref{with-absorber}, we find that for a directional antenna, the metasurface can still contribute similar capacity improvements to without multipath. However, the results from omni-directional antennas are different -- when the transmitted power is lower than $2$~mW, the metasurface will no longer enhance the passing signal, and in fact degrade the channel capacity.
Directional antennas ``concentrate'' the signals through the metasurface, so the incident power is higher and the metasurface can let more power through; omni-directional antennas do not send as much incident power to the metasurface, so the enhancement from polarization matching may not compensate for the loss through the surface. For that reason, the metasurface can effectively block out some weaker multipath components, and therefore the endpoint receiver will not get constructive interference from these multipath components. In that sense, it is possible the metasurface can reduce performance.
%The reason is the gain of the omni-directional antenna~($6$~dBi) is lower than that of the directional antenna~($10$~dBi); when transmit power is too low, the loss incurred by the metasurface will be greater than the amount of improvement; also when the metasurface is not present, the multipath reflections introduced by the environment cause the received signal to be stronger than with no multipath component.

%\noindent\textbf{Incident power at the metasurface.}
%As is the case with previous environment-based approaches~\cite{li2019towards, nsdi2020RFous}, the amount of signal quality change that can be effected by \systemname depends on the incident power at the metasurface. Since we cannot accurately measure this, Figure~\ref{without-absorber}~(a) reported the transmit power from the endpoints placed at the same location as a proxy for the incident power. This experiment suggests that \systemname may perform poorly for BLE \textit{transmitters}. Nevertheless, we believe \systemname could still help Bluetooth \textit{receivers} when the transmitter is a higher-power device, such as a mobile handset.

\subsubsection{Experiments with Low-cost IoT Devices}
Finally, we evaluate \systemname's performance with low-cost IoT devices. We perform tests with conventional Wi-Fi  and Bluetooth links. The Wi-Fi link is between a Wi-Fi router and an Arduino with a low-cost ESP8266 module, and the Bluetooth link is between a Huawei Watch and a Raspberry Pi 3. From the signal power distributions shown in Figure~\ref{IoT-devices}, we find that for the Wi-Fi link, \systemname creates around $10$~dB signal power improvement in a mismatched polarization setup, which looks similar to the matched configuration depicted in Figure~\ref{motivation}~(a). While the improvement for the Bluetooth link is lower, this is expected according to the result shown in Figure~\ref{without-absorber}~(a), which indicates the amount of signal quality change that can be affected by \systemname depends on the incident power level at the metasurface. Nevertheless, we believe \systemname could still help Bluetooth \textit{receivers} when the transmitter is a higher-power device, such as a mobile handset.
%This further validates the effectiveness of \systemname. We believe \systemname also has great potential to enhance other low-cost IoT communication links such as Zigbee or Bluetooth Low Energy.

We next study how \systemname performs when operating across a wider band channel. Most Wi-Fi transmissions today leverage OFDM over $20-40$~MHz. To answer this question, we leverage GIGABYTE mini-PCs equipped with conventional off-the-shelf Intel 5300 wireless cards as transceivers to conduct an experiment -- the center frequency and bandwidth are $2.47$~GHz and $20$~MHz, respectively. Figure~\ref{OFDM} shows the 
channel gain measurements with and without the metasurface in a mismatched configuration. The results show that the overall channel gain is improved, but the enhancement of individual subcarriers are different given the specific bias voltages used for the metasurface; this is consistent with the simulation results shown in Figure~\ref{polarization_results}. 
The subcarrier channel gains with the metasurface are more consistent over frequency than without. We believe the presence of \systemname blocks weak multipath signal components that traverse longer paths and tend to exhibit frequency-selectivity more, and the remaining components are aligned through polarization rotation.

\begin{figure*}[!t]
  \centering
  %\vspace{-0.1cm}
  \subfigure[24cm Tx-Metasurface distance.]{
  \includegraphics[width=0.23\textwidth]{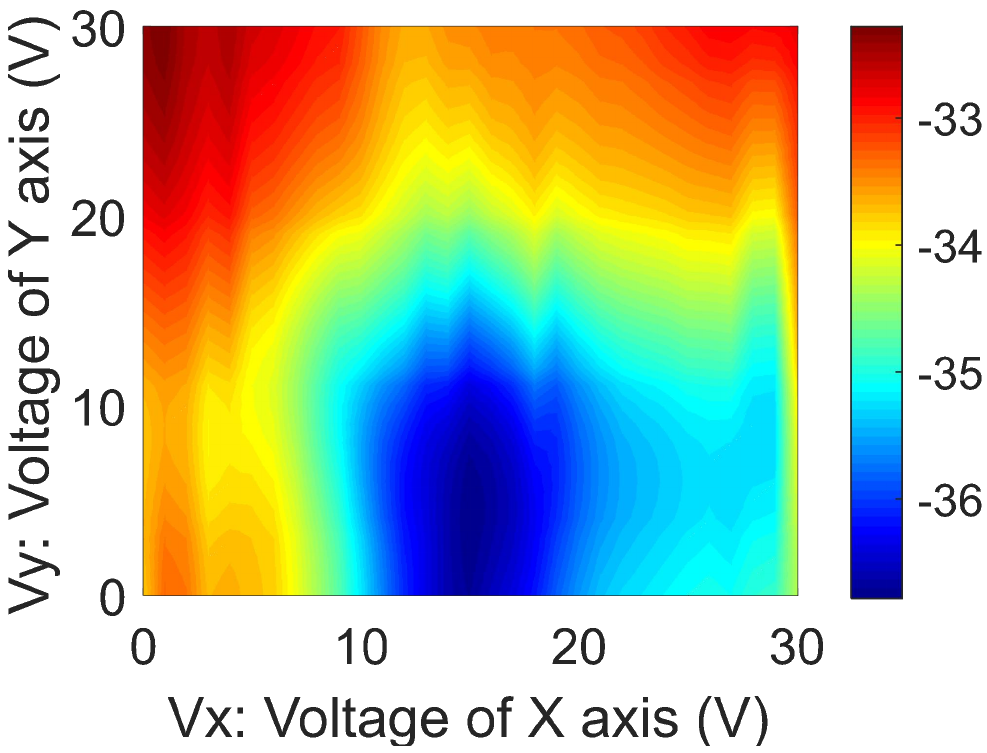}}
  \hspace{0.15cm}
  \subfigure[30cm Tx-Metasurface distance.]{
  \includegraphics[width=0.23\textwidth]{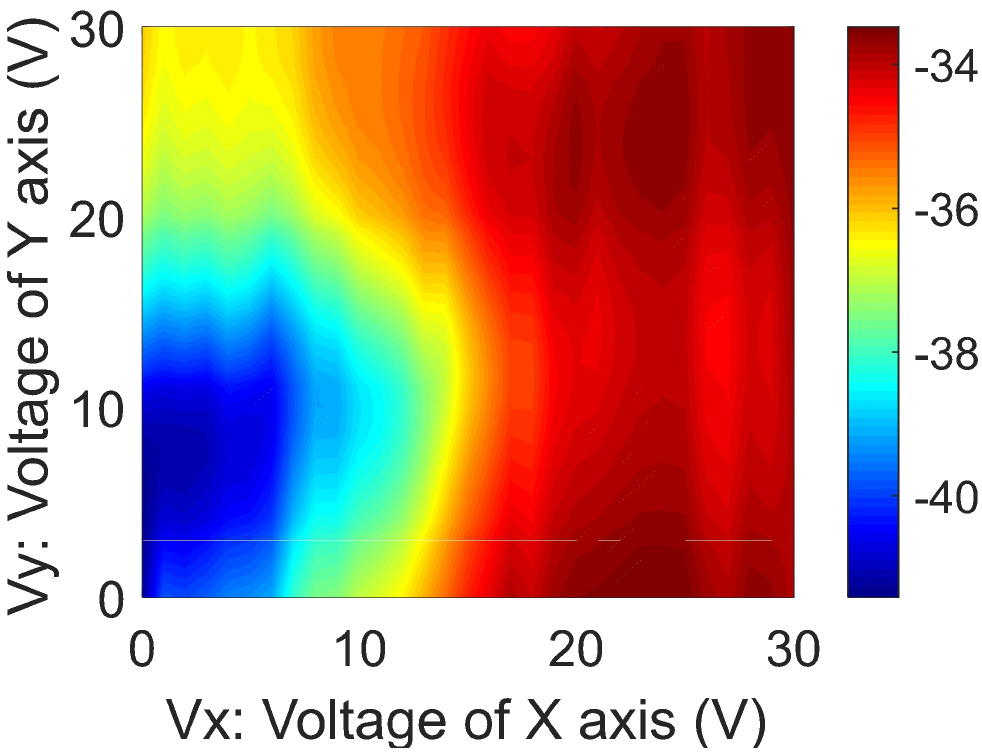}}
  \hspace{0.15cm}
  \subfigure[36cm Tx-Metasurface distance.]{
  \includegraphics[width=0.23\textwidth]{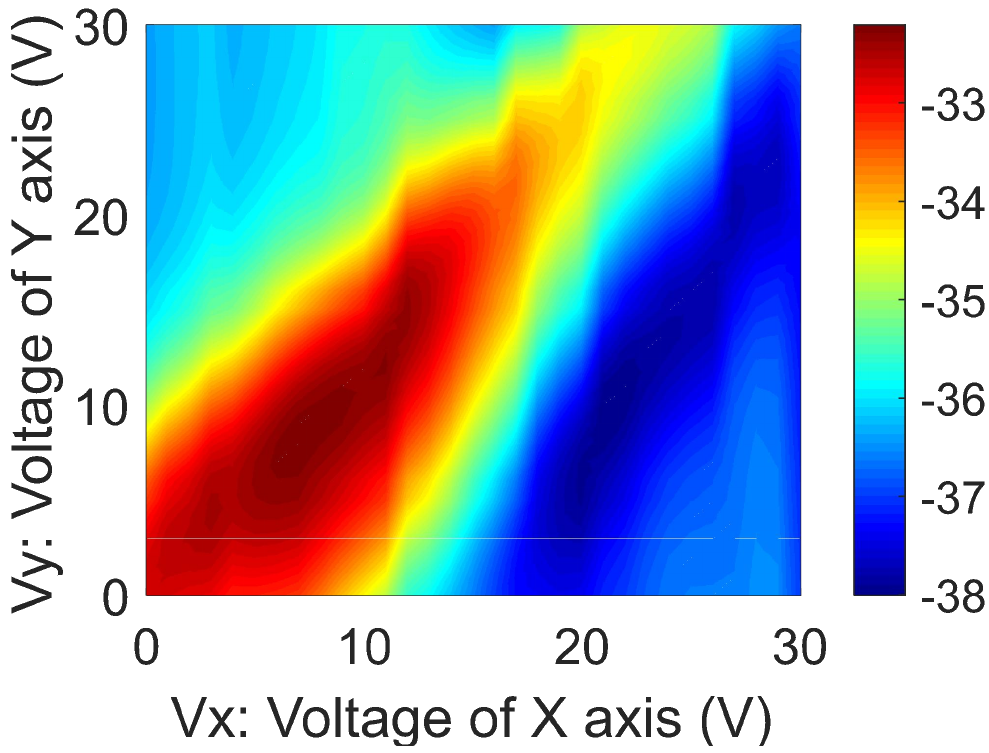}}
  \hspace{0.15cm}
  \subfigure[42cm Tx-Metasurface distance.]{
  \includegraphics[width=0.23\textwidth]{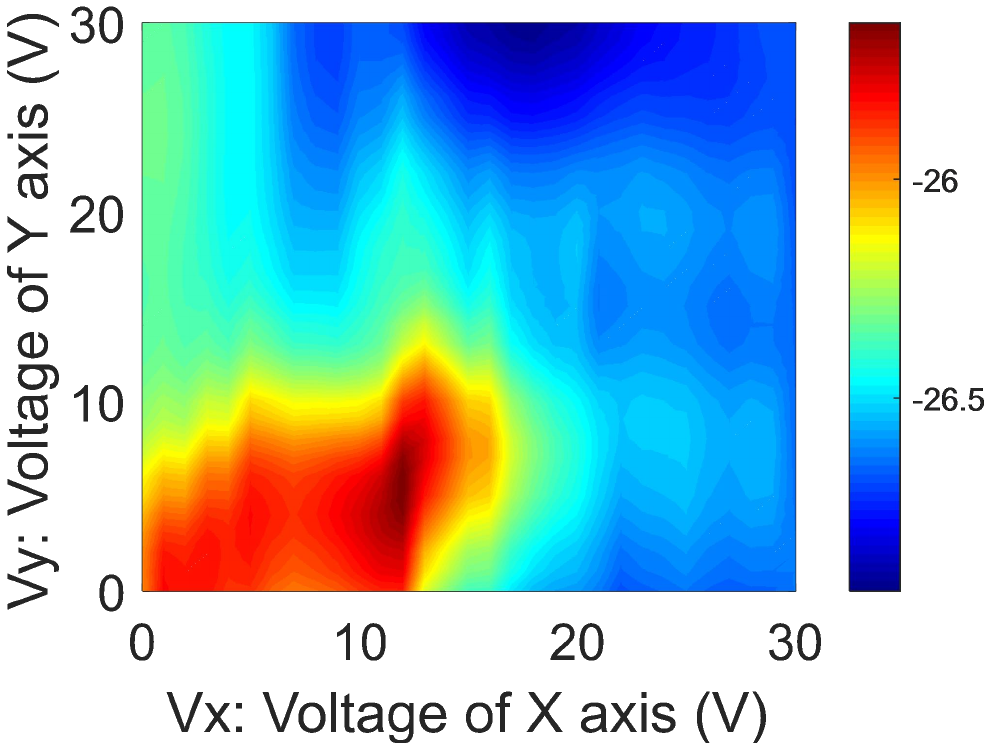}}
  \hspace{0.15cm}
  \subfigure[48cm Tx-Metasurface distance.]{
  \includegraphics[width=0.23\textwidth]{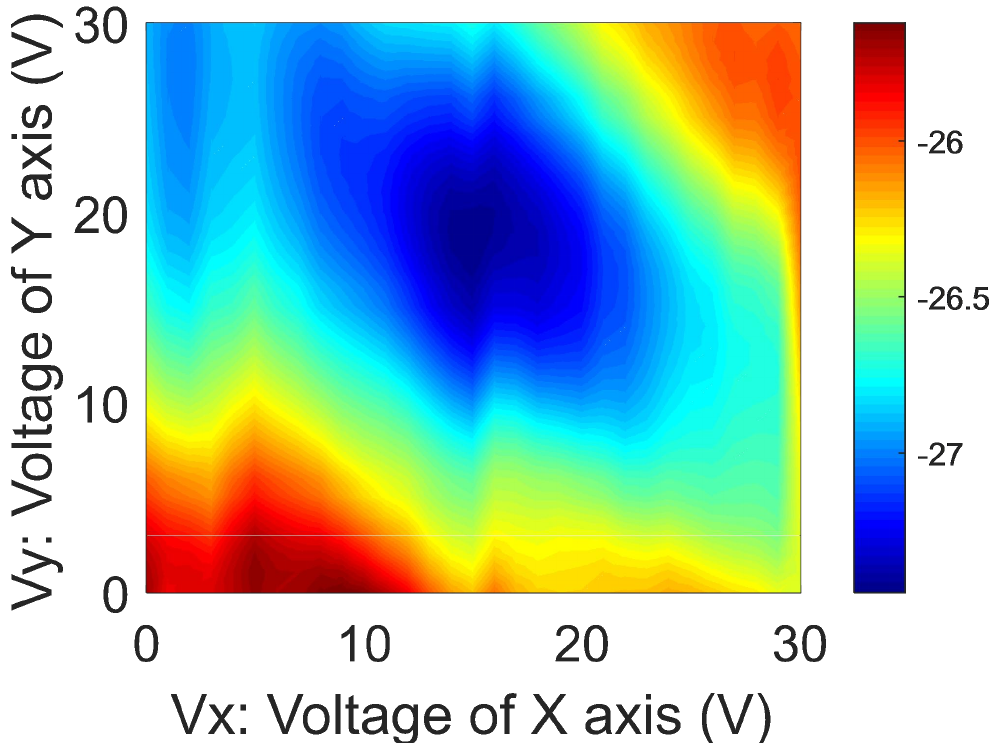}}
  \hspace{0.15cm}
  \subfigure[54cm Tx-Metasurface distance.]{
  \includegraphics[width=0.23\textwidth]{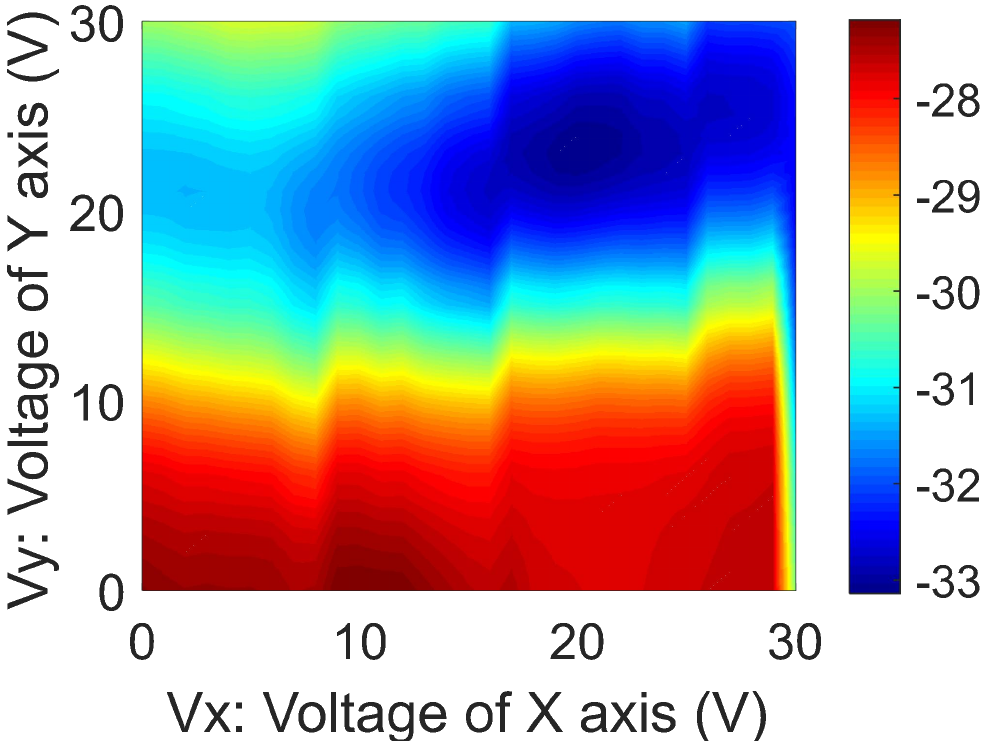}}
  \hspace{0.15cm}
  \subfigure[60cm Tx-Metasurface distance.]{
  \includegraphics[width=0.23\textwidth]{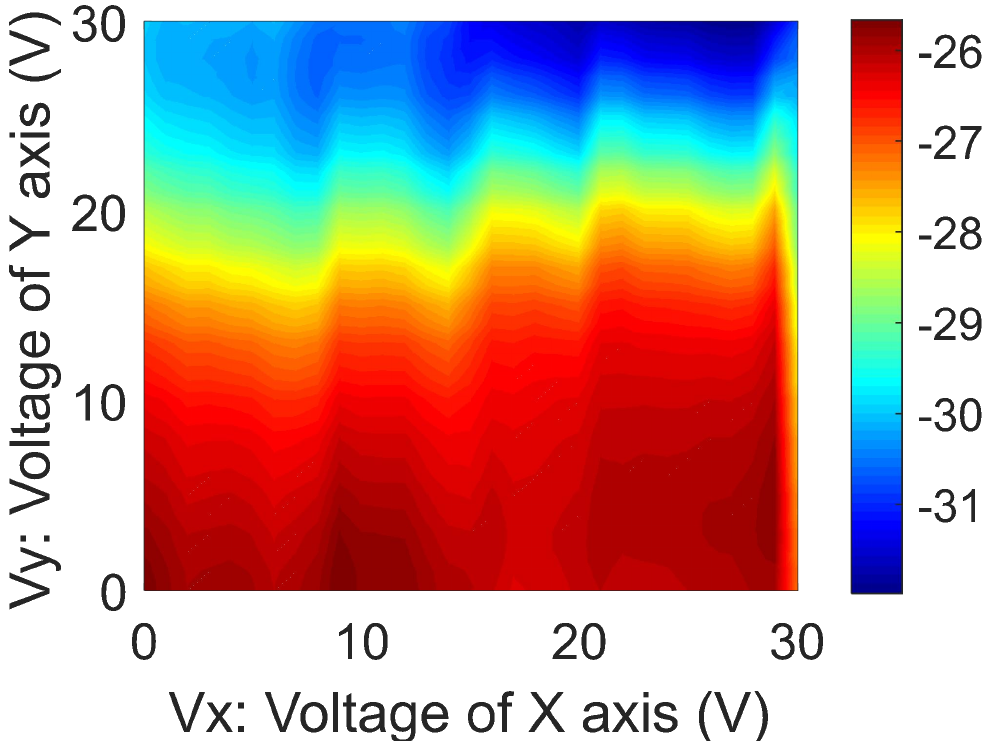}}
  \hspace{0.15cm}
 \subfigure[66cm Tx-Metasurface distance.]{
  \includegraphics[width=0.23\textwidth]{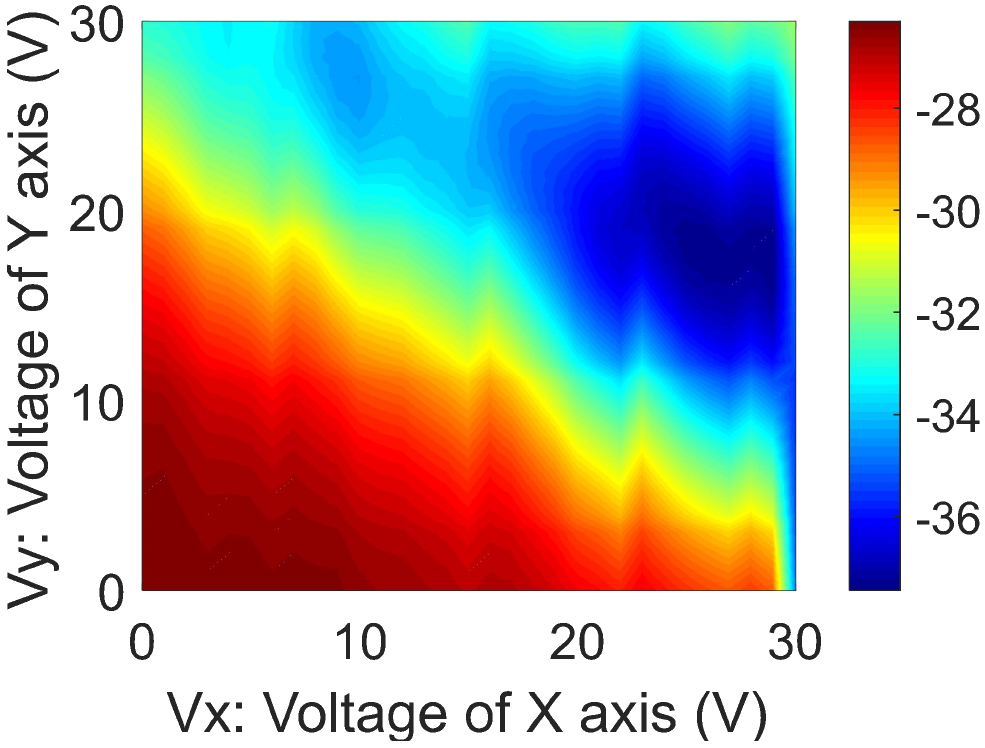}}
  %\vspace{-0.2cm}
  \caption{Experimental results in reflection scenarios. We find that \systemname also changes the reflective signal power.} 
  \label{fig:reflection-heatmaps}
\end{figure*}

\begin{figure*}[!t]
   \centering
   %\vspace{-0.1cm}
  \begin{minipage}[c]{0.47\textwidth}
  \includegraphics[width=1\textwidth]{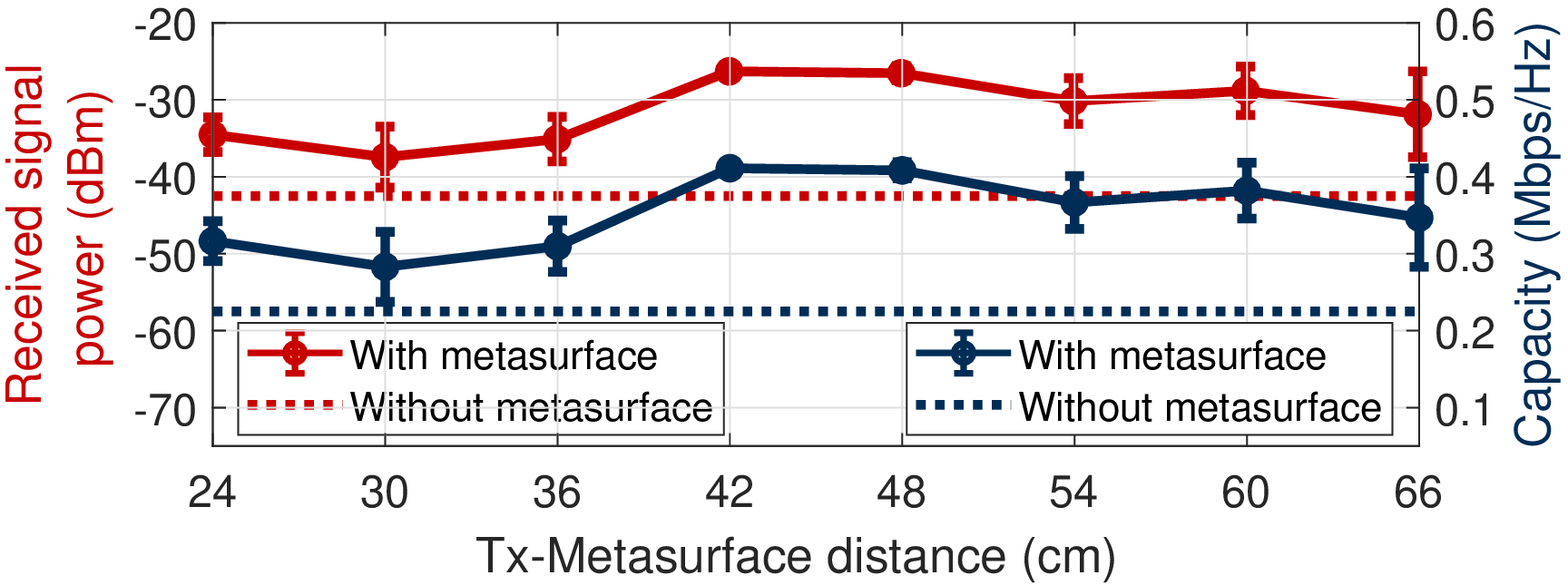}
  %\vspace{-0.2cm}
  \caption{\systemname provides improvements to channel capacity and power in a reflective configuration. }
  \label{reflection}
  \end{minipage}
  \hspace{0.6cm}
  \begin{minipage}[c]{0.47\textwidth}
  \includegraphics[width=1\textwidth]{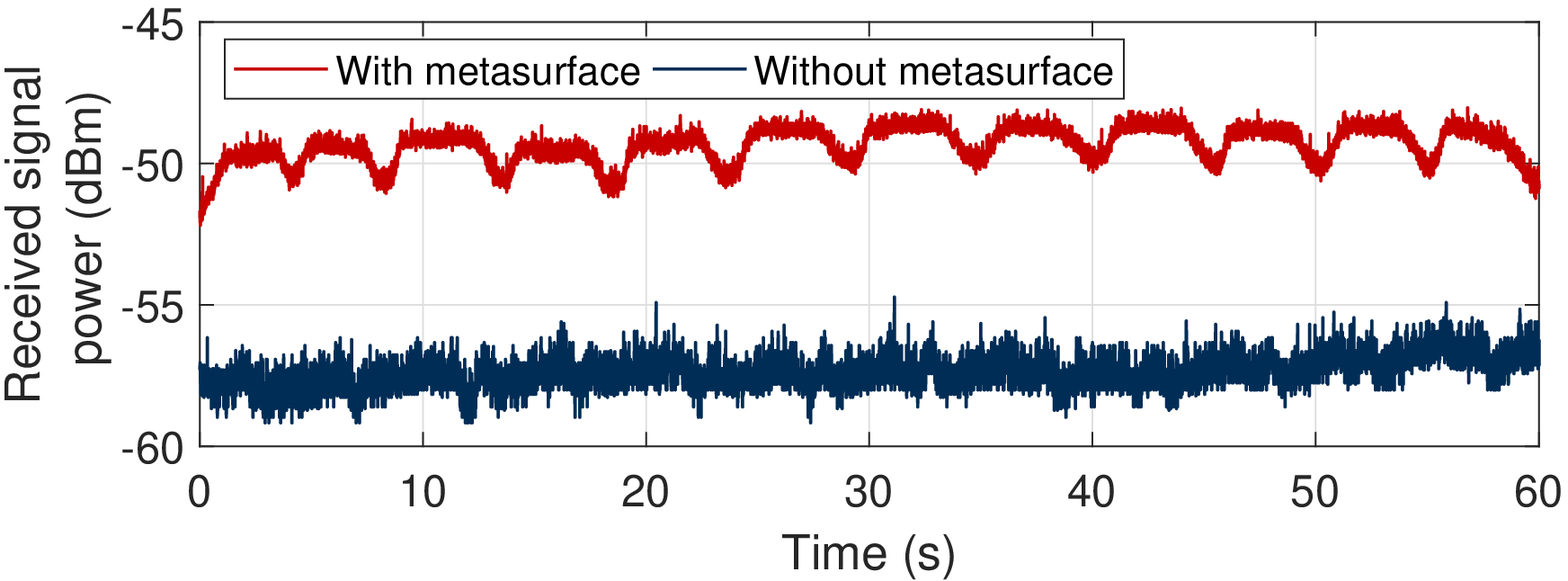}
  %\vspace{-0.2cm}
  \caption{Human respiration sensing results with/without metasurface at low transmitting power of $5$~mW.}
  \label{respiration}
  \end{minipage}
  \vspace{-0.2cm}
\end{figure*}

%\vspace{-0.1cm}
\subsection{Reflective Operation}
\subsubsection{Reflective Signal Enhancement}
%\noindent\textbf{Reflective signal enhancement.}
In addition to evaluating transmissive configurations, we also look at \systemname's effect on reflected signals.
We place the transmitter and receiver on the same side of the metasurface, and separate the transmitter and receiver by $70$~cm. We perform experiments at different Tx-metasurface distances by moving the metasurface along the vertical line of the transceiver pair. Figure~\ref{reflection} shows the maximum received signal power and channel capacity with the metasurface , as well as the baseline measurements without the metasurface in a mismatched configuration. These results show that \systemname also has a positive impact on the reflection scenario---the signal power and capacity can be improved with respect to mismatch by as much as $17$~dB and $180$~kbps/Hz, respectively. However, the signal power difference over voltage combinations~(see Figure~\ref{fig:reflection-heatmaps}) is much smaller than that in the transmission scenario. We believe this is because the rotation will be cancelled after reflection. 

\subsubsection{Employing \systemname for Sensing}\label{evaluation: sensing}
%\noindent\textbf{Employing \systemname for sensing.}
Based on the reflective configuration, we conjecture that \systemname can be utilized to enhance sensing. To validate this, we consider human respiration detection as a case study to test \systemname's potential. In this experiment, the metasurface is placed $2$~m away from the center of the transceiver pair, the human subject is located on the side between the transmitter and the metasurface. First, we remove the metasurface and reduce the transmitting power to where the human subject's respiration can no longer be detected from the received signal. Then we introduce the
metasurface at the predetermined location, and measure the received signal strength. The detection results with and without the metasurface are plotted in Figure~\ref{respiration}. It is clear that the metasurface can enhance the reflected signal and allow the target's respiration rate to be detectable under a low transmit power configuration ($5$~mW). We believe that \systemname can also be extended to other low SNR sensing applications~\cite{zhao2016emotion, ha2020food}.

%\noindent\textbf{\blue{Arm motion detection.}}

%\noindent\textbf{\blue{Goods direction check.}}

\section{Discussion and Future Work}

\noindent\textbf{Scaling to a dense IoT deployment.}
 This work marks the first step towards mitigating polarization issues for individual communication links with a \systemname prototype. Next, we plan to scale up the size of the metasurface for a larger deployment and explore more challenging multi-link scenarios. When there are multiple IoT devices in different polarization orientations, tuning the signal polarization can lead to a new form of polarization reuse or access control and improve the network throughput of a dense IoT deployment.

\noindent\textbf{Adapting to device mobility.} 
The current search time for optimal voltage is limited by the switching speed of the commodity power supply we used, hence there is still a latency issue in mobility scenarios. In the future, we will look at methods that can speed this up once the relative antenna orientations are determined and then track the changes.

% \noindent\textbf{Incident power at the metasurface.}
% As is the case with previous environment-based approaches~\cite{li2019towards, nsdi2020RFous}, the amount of signal quality change that can be effected by \systemname depends on the incident power level at the metasurface. Since we cannot accurately measure this incident power, Figure~\ref{without-absorber}~(a) reported the transmit power from the endpoints placed at the same location as a proxy for the incident power. This experiment suggests that \systemname might perform poorly for low-power Bluetooth \textit{transmitters}. Nevertheless, we believe \systemname could still help Bluetooth \textit{receivers} when the transmitter is a higher-power device, such as a mobile handset.

\section{Related Work}
Broadly speaking, our work is related to work in three areas:

\noindent\textbf{Endpoint optimizations.}
Most efforts for improving communication quality focuses on controlling the endpoints themselves. For instance, Multiple Input, Multiple Output (MIMO) links leverage multiple antennas to exploit spatial diversity at a sub-link level, while Multi-User MIMO exploits spatial diversity at an inter-link level~\cite{hamed2016real, hamed2018chorus}. Massive MIMO introduces many more antennas at an access point than both radio chains and users, so that the AP may search for a set of antennas that forms a well-conditioned MIMO channel to those users~\cite{shepard2012argos, rahul2012jmb, yang2013bigstation}. However, these approaches are fundamentally limited if the cause of performance loss is antenna polarization mismatch between endpoints. %for two reasons. On one hand, for a given wireless channel, there are limited degrees of freedom the endpoints can use to increase throughput and mitigate interference. On the other hand, approaches such as phased arrays and distributed antenna systems rely on bulky and/or expensive transmission
%lines (\emph{i.e.}, coaxial cables, optical fiber, or waveguides) to distribute the signal among antennas.
At the endpoints, the only directly relevant mitigation strategies are to use either circularly polarized antennas or multiple linearly polarized antennas. Once the antennas are fixed, not much can be done about polarization match at the endpoints. Using an antenna array like massive MIMO can enhance the received signal power, but without directly addressing the polarization issue. 

In contrast, an approach that changes the radio environment itself~(e.g., deploying low-cost reflectors) offers the possibilities to increase the number of degrees of freedom.
%which can be achieved by combining the wireless propagation environment with low-cost reflectors~\cite{tan2016increasing, abari2017enabling, li2019towards, nsdi2020RFous}.

\noindent\textbf{Environment-based optimizations.} 
Previous work on radio environment optimizations fall into two categories: phase-based and amplitude-based. Initial attempts of phase based approaches such as leveraging static mirrors~\cite{zhou2012mirror} or programmable phased-array reflectors~\cite{abari2016cutting, abari2017enabling, tan2016increasing} are in the ability of generating constructive propagation paths. These methods focus on millimeter wave links on high frequency bands~(\emph{i.e.}, $10$~GHz and above). More generally, several proposals argue for dynamically reconfiguring the radio environments~\cite{welkie2017programmable, ris-access2019, smartenv-2019, bjrnson2019massive, poor-2020lisa, irs-commsmag2020, visorsurf-commsmag2018}.
Specifically, recent prototypes manipulate the signal propagation behavior in the 2.4~GHz band, by using a large array of inexpensive antennas~\cite{welkie2017programmable, li2019towards, dunna2020scattermimo} or conductive surfaces~\cite{ chan2018surface} as phase shifting elements. These systems align phase elements according to a channel decomposition. Amplitude-based designs, such as RFocus \cite{nsdi2020RFous}, sidestep the difficulty in measuring phase. Based on the signal amplitude measurements from the receiver, RFocus configures the signal to either pass through or reflect from the surface element by setting the “on” or “off” state of each element, so that the transmitted signal is focused at the intended receiver.

Orthogonal to prior work that aligns multiple paths to achieve beamforming effects or improve spatial multiplexing efficiency, \systemname optimizes low-cost IoT communication links by specifically overcoming the pervasive issue of polarization mismatch that affects both single and multi-path communication.

%\blue{Unlike prior work that aligns multiple paths to achieve beamforming effects or improve spatial multiplexing efficiency, \systemname optimizes low-cost IoT communication links by overcoming the pervasive issue of polarization mismatch that affects both single and multi-path communication.}

\noindent\textbf{Metamaterials.}
Metamaterials are an earlier, more general form of metasurfaces that are constructed in 3D rather than 2D. These are artificially constructed with special properties. Recent work in the applied physics community has developed metamaterials that can directly alter
existing signals in the environment itself, such as creating materials with a negative refraction index~\cite{kourakis2005nonlinear} and engineering complex beam patterns~\cite{orlov2011engineered}. Other work has verified the feasibility of leveraging metamaterials to change the signal polarization~\cite{hao2007manipulating, zheng2017ultra, yang2016programmable, wu2019tunable, zhang2020polarization}.  With a biasing network, different voltages are provided to 
diodes integrated on the metasurface for rotating the polarization of a transmitted wave. While these designs have shown great promise in controlled experiments that quantify performance in a higher frequency band~(\emph{i.e.}, $>5$~GHz), they were constructed using expensive, low loss substrate materials such as Rogers or F4B. Furthermore, they have not been integrated into an end-to-end system that optimizes signal paths in real time.

In contrast, we present an end-to-end system incorporating the structure of a metasurface design for the $2.4$~GHz ISM band using cheaper, but higher loss FR4 material, and specifically control the structure's polarization rotation to optimize the communication link between a pair of devices.

\section{Conclusion}
This paper highlights the under-appreciated issue of polarization mismatch for low-cost IoT devices that are physically limited to employing a single low-quality antenna. We present \systemname, a system designed to mitigate the polarization mismatch without hardware modifications to the endpoints.
\systemname is capable of manipulating the polarization state of the signal arriving at the receiver with a tunable metasurface structure made with cheap material. %to avoid the polarization mismatch between the transmitter and the receiver. Technically, \systemname leverages a metasurface structure made with cheap material to act as a polarization rotator. \systemname has been successfully integrated with an IoT network.
It can optimize the communication quality in real time, and enhance the performance of sensing applications. %\blue{The focus of this paper is on improvements at the physical layer, we also plan to explore new challenges in the link layer and the MAC layer in the future.}
%This work marks the first step towards mitigating polarization issues for individual communication links with a \systemname prototype. Next, we plan to scale up the size of the metasurface for a larger scale deployment and explore more challenging multi-link scenarios. When there are multiple IoT devices in different polarization orientations, tuning the signal polarization can lead to a new form of polarization reuse or access control and improve the network throughput for dense IoT deployments.
\section*{Acknowledgement}
We thank our shepherd Fadel Adib, and the anonymous reviewers for their helpful feedback in improving the paper. This work is supported by the National Science Foundation under Grant Nos. CNS-1617161, CNS-1763212 and
CNS-1763309, the National Natural Science Foundation of China under Grant Nos. 61672428 and 62061146001, the ShaanXi Science and Technology Innovation Team Support Project under Grant No. 2018TD-026. Any opinions, findings, and conclusions or
recommendations expressed in this material are those of the
author(s) and do not necessarily reflect the views of the National Science Foundation.
%-------------------------------------------------------------------------------

\clearpage
\pagenumbering{roman}

%-------------------------------------------------------------------------------
\let\oldbibliography\thebibliography
\renewcommand{\thebibliography}[1]{%
  \oldbibliography{#1}%
  \setlength{\parskip}{0pt}%
  \setlength{\itemsep}{0pt}%
}
\bibliographystyle{concise2}
%\bibliography{\jobname}
\begin{raggedright}
\bibliography{refs}
\end{raggedright}
%%%%%%%%%%%%%%%%%%%%%%%%%%%%%%%%%%%%%%%%%%%%%%%%%%%%%%%%%%%%%%%%%%%%%%%%%%%%%%%%
\end{document}